\documentclass[12pt,preprint]{aastex6}
\usepackage{amsmath}
\usepackage{graphicx}
\usepackage{enumitem}
\usepackage{multirow}
\usepackage{color}

\shorttitle{Transverse Wave Induced Kelvin-Helmholtz Rolls in Spicules}
\shortauthors{P. Antolin et al.}

\begin{document}

\title{Transverse Wave Induced Kelvin-Helmholtz Rolls in Spicules}

\author{P. Antolin\altaffilmark{1}, D. Schmit\altaffilmark{2,3}, T. M. D. Pereira\altaffilmark{4,5}, B. De Pontieu\altaffilmark{2,4,5}, I. De Moortel\altaffilmark{1}}
\affil{\altaffilmark{1}School of Mathematics and Statistics, University of St. Andrews, St. Andrews, Fife KY16 9SS, UK\\
\altaffilmark{2}Lockheed Martin Solar and Astrophysics Laboratory, B/252, 3251 Hanover Street, Palo Alto, CA 94304, USA\\
\altaffilmark{3}Now at: Catholic University, Department of Physics, 620 Michigan Ave., N.E., Washington, DC 20064\\
\altaffilmark{4}Rosseland Centre for Solar Physics, University of Oslo, P.O. Box 1029 Blindern, N-0315 Oslo, Norway\\
\altaffilmark{5}Institute of Theoretical Astrophysics, University of Oslo, P.O. Box 1029 Blindern, N-0315 Oslo, Norway
}
\email{patrick.antolin@st-andrews.ac.uk}

\begin{abstract}

In addition to their jet-like dynamic behaviour, spicules usually exhibit strong transverse speeds, multi-stranded structure and heating from chromospheric to transition region temperatures. In this work we first analyse \textit{Hinode} \& \textit{IRIS} observations of spicules and find different behaviours in terms of their Doppler velocity evolution and collective motion of their sub-structure. Some have a Doppler shift sign change that is rather fixed along the spicule axis, and lack coherence in the oscillatory motion of strand-like structure, matching rotation models or long wavelength torsional Alfv\'en waves. Others exhibit a Doppler shift sign change at maximum displacement and coherent motion of their strands, suggesting a collective MHD wave. By comparing with an idealised 3-D MHD simulation combined with radiative transfer modelling, we analyse the role of transverse MHD waves and associated instabilities in spicule-like features. We find that Transverse Wave Induced Kelvin-Helmholtz (TWIKH) rolls lead to coherence of strand-like structure in imaging and spectral maps, as seen in some observations. The rapid transverse dynamics and the density and temperature gradients at the spicule boundary lead to ring-shaped \ion{Mg}{2} k and \ion{Ca}{2} H source functions in the transverse cross-section, potentially allowing IRIS to capture the KHI dynamics. Twists and currents propagate along the spicule at Alfv\'enic speeds, and the temperature variations within TWIKH rolls produce sudden appearance / disappearance of strands seen in Doppler velocity and in \ion{Ca}{2} H intensity. However, only a mild intensity increase in higher temperature lines is obtained, suggesting there is an additional heating mechanism at work in spicules.

\end{abstract}


\section{Introduction}

Spicules are chromospheric jets protruding into the corona, ubiquitously found on the Sun. Properties such as dynamics, morphology and heating, suggest that these jets may play a substantial role as energy conduits to the corona \citep{DePontieu_etal_2011Sci...331...55D}. 

Two types of spicules have been proposed \citep{DePontieu_etal_2007PASJ...59S.655D}, both of which exhibit strong transverse (combining swaying \& torsional) motions \citep{Sekse_2013ApJ...769...44S, DePontieu_2014Sci...346D.315D, Pereira_2012ApJ...759...18P, Pereira_2016ApJ...824...65P}. Type I spicules typically show full parabolic paths in chromospheric lines, with velocities on the order of $15-40$~km~s$^{-1}$ and lifetimes of about $3-10$~min \citep{Pereira_2012ApJ...759...18P}. Type II spicules differ from their type I counterparts in terms of the faster observed longitudinal speeds ($30-100$~km~s$^{-1}$), their fast disappearance in the \ion{Ca}{2}~H passband but increase in visibility in higher temperature emission lines \citep{DePontieu_2009ApJ...701L...1D,Pereira_2014ApJ...792L..15P, Rouppe_2015ApJ...799L...3R}, and their multi-stranded structure \citep{Suematsu_2008ASPC..397...27S,Skogsrud_2014ApJ...795L..23S}. 

In high resolution tunable Fabry-P\'{e}rot instruments such as \textit{SST}/CRISP, fast disappearance of the entire structure for type II spicules in high blue and red-shift positions is observed, suggesting very high, even non-physical propagation speeds \citep{Pereira_2016ApJ...824...65P, Kuridze_2015ApJ...802...26K,Shetye_2017arXiv170310968S}. This fact has led to consider different interpretations for spicule formation, such as the combination of different line-of-sights and the expected curved shape of chromospheric current sheets within flux tubes \citep{Judge_2012ApJ...755L..11J}. This sudden appearance and disappearance in narrow passbands at high blueshifts and redshifts, observed throughout the solar disk and off-limb, cannot be explained by a simple geometric expansion of the loop combined with very fast longitudinal speeds.  \citep{Sekse_2013ApJ...769...44S,Pereira_2016ApJ...824...65P}. The observed transverse swaying and torsional component in spicules (of the order of $25-30$~km~s$^{-1}$) was proposed as an explanation for this effect \citep{DePontieu_2012ApJ...752L..12D}, a motion that could be caused by a mix of torsional Alfv\'en and kink-mode waves. This sudden appearance and disappearance in narrow passbands produces contiguous strand-like structure of spicules in Doppler maps, with opposite velocities along their entire lengths, and with a high frequency oscillatory change in the Doppler signal. The torsional Alfv\'en wave interpretation for this effect was recently supported by \citet{Srivastava_2017NatSR...743147S}. The latter work lacks, however, any forward modelling, and a clear link between the numerical modelling cannot be established with the observations.

While general consensus exists for the generation mechanism of type I spicules being longitudinal slow modes steepening into shocks \citep{DePontieu_2005ApJ...624L..61D}, there is no work that fully reproduces type II spicules. The 3-D MHD models of \citet{Martinez-Sykora_2013ApJ...771...66M} and \citet{Iijima_2016PhDT.........5I}, encompassing both the sub-photospheric region and corona, manage to generate spicules with significant transverse displacements (and rotational components) from reconnection at photospheric heights combined with a slingshot effect by the Lorentz force. However, little strand-like structure in intensity and Doppler is obtained, and the characteristic disappearance in the \ion{Ca}{2}~H band is not achieved. On the other hand, the recent model by \citet{Martinez-Sykora_2017} shows that ambipolar diffusion may be the key to the generation of type II spicules. The addition of this mechanism increases the magnitude of the Lorentz force and the ensuing slingshot effect, thereby non-linearly generating a strong enough slow mode shock that pushes the mass upward (the spicule). Accompanying strong currents and thermal fronts are produced from the Alfv\'enic perturbations, which lead to heating to transition region and coronal temperatures (and disappearance of \ion{Ca}{2}~H emission) during the lifetime of the spicule \citep{DePontieu_2017ApJ...845L..18D}, and to fast apparent propagation \citep{DePontieu_2017ApJ...849L...7D}, potentially explaining the observed fast plane-of-the-sky (POS) velocities in transition region lines \citep{Tian_2014Sci...346A.315T,Narang_2016SoPh..291.1129N}. However, the model is limited to 2.5-D MHD with low spatial resolution. Hence, the dynamic instabilities and turbulence associated with transverse MHD waves and resonant absorption, which are expected to play a role in the spicule morphology and thermodynamic evolution, are not captured. It is therefore unclear how much of the observed type II spicules characteristics can be reproduced with this model, particularly concerning the observed variability in the Doppler signal. 

Transverse MHD waves are observed in the solar atmosphere \citep{DePontieu_2007Sci...318.1574D,Tomczyk_2007Sci...317.1192T,Arregui_2012LRSP....9....2A,DeMoortel_Nakariakov_2012RSPTA.370.3193D} and constitute one of the prime coronal heating candidates, thanks to their ability to carry large amounts of energy into the corona \citep[see e.g. review by][]{Arregui_2015RSPTA.37340261A}. Many of these energy estimates (and therefore much of the heating candidate status) are actually based on observations of these waves in the chromosphere, and particularly along spicules \citep{DePontieu_2007Sci...318.1574D,He_2009ApJ...705L.217H,Jess_2012ApJ...744L...5J,Kuridze_2016ApJ...830..133K,Pereira_2012ApJ...759...18P,Pereira_2016ApJ...824...65P}, where large amplitudes of tens of km s$^{-1}$ are usually observed. Such transverse, wave-like motions of spicules constitute a key observed property of these phenomena. 

Transverse MHD waves can easily trigger the Kelvin-Helmholtz Instability (KHI) \citep{Terradas_2008ApJ...687L.115T,Antolin_2014ApJ...787L..22A,Magyar_2016ApJ...823...82M}, with vortices excited mostly in the transverse plane to the loop axis. These 3-D vortices are known as Transverse Wave Induced K-H rolls (TWIKH rolls), and generate a turbulent-like regime in which energy can be dissipated through viscosity and resistivity \citep{Magyar_2016ApJ...823...82M, Howson_2017AA...602A..74H}. Recently, a new turbulent regime from unidirectional Alfv\'enic waves has been discovered, which relaxes the condition of having counter-propagating waves for the onset of turbulence \citep{Magyar_2017arXiv170202346M}. Although the KHI is produced preferentially in the absence of an inhomogeneous boundary due to the short spatial scales of the shear \citep{Terradas_2008ApJ...687L.115T}, the presence of an inhomogeneous layer can maintain the KHI for longer times. Indeed, the resonant absorption mechanism associated with transverse MHD waves  continuous to act until the global kink mode is fully damped \citep{Goossens_2002AA...394L..39G}.  A continuous driving of these waves, as investigated by \citep{Karampelas_2017AA...604A.130K}, is observed to completely deform the original flux tubes and leads to heating preferentially by Ohmic dissipation towards the footpoints. The fast longitudinal speeds observed in type II spicules also suggest that the KHI wave vectors have an important longitudinal component \citep{Zhelyazkov_2015AdSpR..56.2727Z,Ajabshirizadeh_2015ApSS.357...33A,Zaqarashvili_2015ApJ...813..123Z}, leading to helical TWIKH rolls that could be responsible for observed vortex-like structure \citep{Kuridze_2016ApJ...830..133K}. 

In this work we analyse and compare spicules observed with \textit{Hinode} \citep{Kosugi_2007SoPh..243....3K} and the \textit{Interface Region Imaging Spectrograph} \citep[IRIS;][]{DePontieu_2014SoPh..289.2733D} with a high resolution 3-D MHD model of a spicule-like feature subject to transverse MHD waves. Our aim is not to model the generation of a spicule, but to study the effect that these waves may have on the spicule during its lifetime, and identify which observational characteristics can be attributed to the presence of these waves. Such kink waves may be produced by the spicule driver \citep{He_2009ApJ...705L.217H,Jess_2012ApJ...744L...5J}, as in the recently proposed model driven by  ambipolar diffusion of \citet{Martinez-Sykora_2017}, which, however, does not include the high-resolution modelling, wave mode coupling and the associated dynamic instabilities we consider here.

\section{Observations}\label{obs}
\subsection{\textit{IRIS \& Hinode} Coordinated Observations}

\begin{figure}[!ht]
\begin{center}
\includegraphics[scale=0.7]{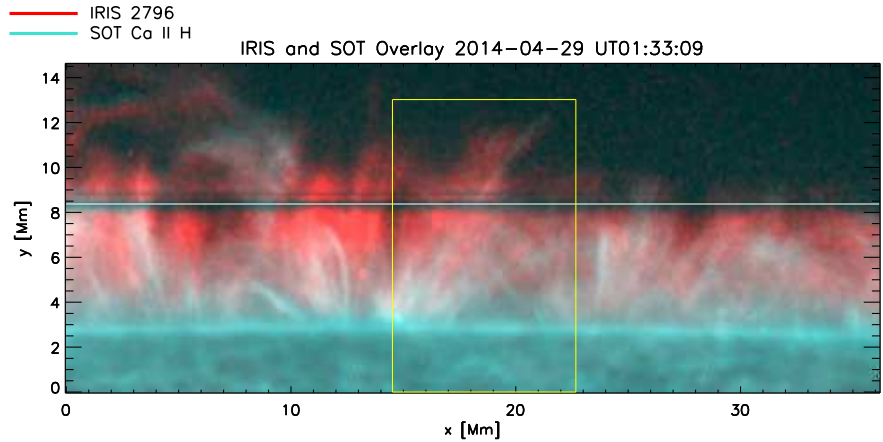}
\caption{\textit{Hinode} and \textit{IRIS} observations of spicules. Composite image of \textit{IRIS} SJI 2796 (red) \& SOT~\ion{Ca}{2}~H line (turquoise). The FOV corresponding to Fig.~\ref{fig2} is marked by the yellow square. The \textit{IRIS} slit location is marked by the horizontal white line. In the animated movie corresponding to this figure the FOVs of the cases shown in Fig.~\ref{fig3} also appear in a yellow square at the time of occurrence. 
\label{fig1}}
\end{center}
\end{figure}

\textit{IRIS} and \textit{Hinode} co-observed at the north limb between 2014 April 29 01:00UT and 02:00UT (IHOP 249). The \textit{IRIS} observing program ran between 2014 Apr 28 22:39UT and Apr 29 02:58 UT, with the \textit{IRIS} slit centered at helioprojective coordinates $(-4,959)$. \textit{IRIS} was rolled $90^\circ$ so that the slit was parallel to the limb and with the middle of the slit $7^{\prime\prime}$ off the limb. The  \textit{IRIS} Slit-Jaw Imager (SJI) and Spectrograph (SG, in sit-and-stare mode) observed with cadences 19~s (SJI) and 9.5~s (SG) with exposure of 8~s (and 1.5~s readout time). The observing program included SJI 1400 and 2796 channels, which provide passband filtered images centred on the emission lines of \ion{Si}{4} at 1402.77~\AA\,~and \ion{Mg}{2}~k~at 2796.35~\AA , respectively. The SJI instrument observed with a field-of-view (FOV) of $120^{\prime\prime}\times120^{\prime\prime}$ and $0.^{\prime\prime}166$~pix$^{-1}$ platescale. The \textit{IRIS} slit is $0.33^{\prime\prime}$ wide. For analysis, we use level~2 data \citep{DePontieu_2014SoPh..289.2733D} and correct for thermal variations of the pointing by co-aligning each image using a cross correlation maximisation routine. We find that the \textit{IRIS} data drifts by $3^{\prime\prime}$ over the 60 minutes of observations overlapping with the SOT observing period. The \textit{IRIS} spectrograph has 25~m\AA~platescale and 53~m\AA~resolution in the NUV. The spectral profiles of the spicules at the slit height exhibit few reversals on average and are fitted as gaussians using a least squares minimisation algorithm. Profile fits with a high $\chi^2$ were discarded.

The \textit{Hinode} Solar Optical Telescope \citep[SOT;][]{Suematsu_2008SoPh..249..197S, Tsuneta_2008SoPh..249..167T} observed between 2014 April 29 01:00UT and 02:00UT with a cadence of 4.8~s (1.8~s exposure). SOT observed a $56^{\prime\prime}\times56^{\prime\prime}$ FOV with $0.^{\prime\prime}11$~pix$^{-1}$ platescale. The entire SOT FOV is included in the \textit{IRIS} FOV. The SOT dataset was processed using the $\tt{FG\_PREP}$ Solarsoft routine, which flat fields and dark subtracts the data. Over the 60 minutes of observations, our dataset drifts approximately $6^{\prime\prime}$ southwest which we correct using a linear offset function.

\begin{figure}[!ht]
\begin{center}
\includegraphics[scale=0.7]{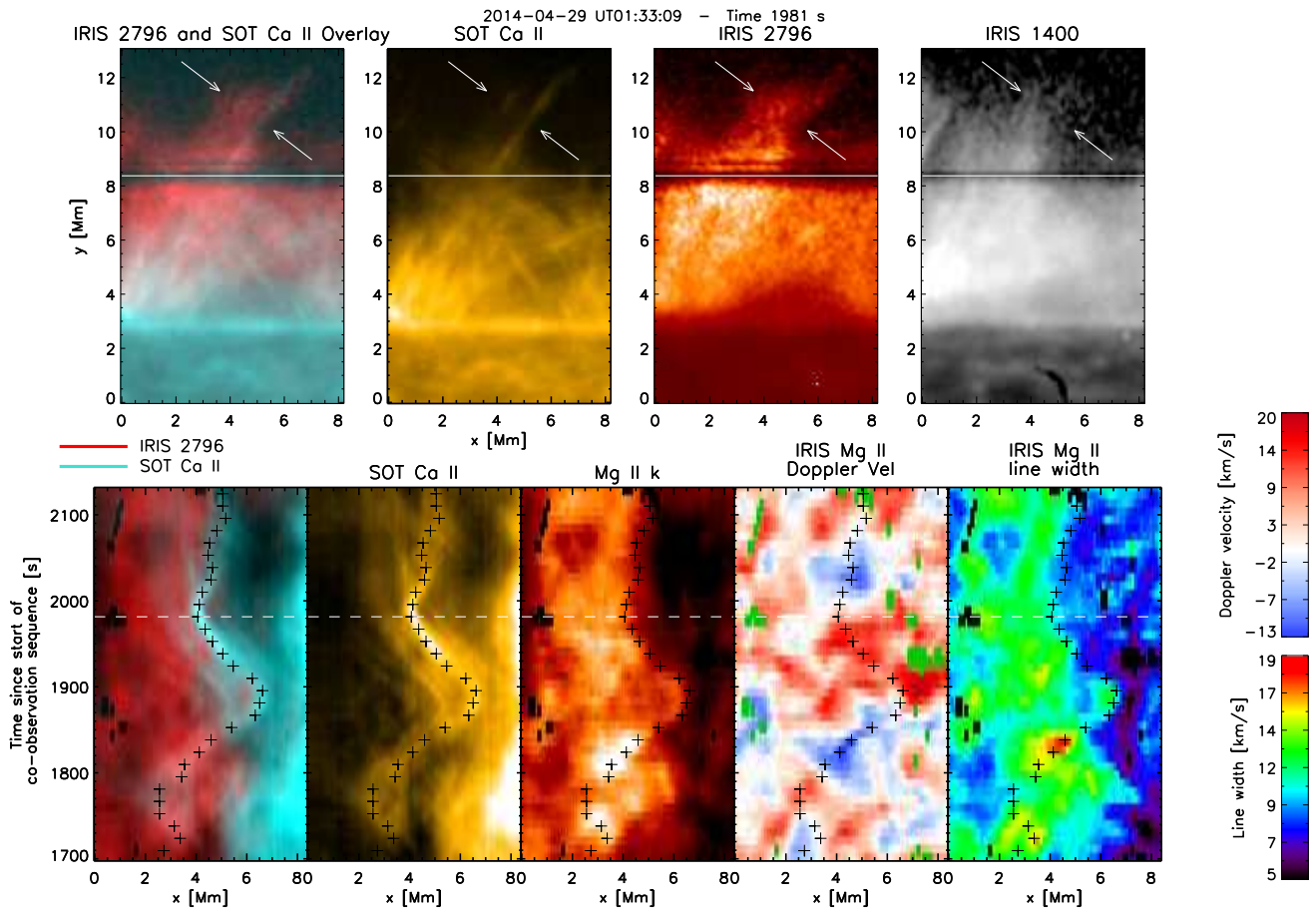}
\caption{\textit{Hinode} and \textit{IRIS} observations of spicules. \textit{top panels:} From left to right, composite image of \textit{IRIS} SJI~2796 (red) \& SOT~\ion{Ca}{2}~H line (turquoise), the corresponding SOT image in the \ion{Ca}{2}~H line (yellow), the SJI~2796 image (red) and the SJI~1400 image (grey). In these images a spicule group is observed (marked within arrows). \textit{bottom panels:} from left to right, time-distance diagrams at the \textit{IRIS} slit location (horizontal white line on top panels) for the composite intensities of \textit{IRIS} SJI~2796 (red) \& SOT~\ion{Ca}{2}~H line (turquoise), the SOT~\ion{Ca}{2}~H line intensity (yellow), the SJI~2796 intensity (red), the Doppler velocity and line width in \ion{Mg}{2}~k. For the SOT intensity diagram we have summed over several cuts perpendicular to the oscillating structure, along 2~Mm length from the location of the \textit{IRIS} slit crossing and above. The black crosses denote locations of maximum \ion{Ca}{2} intensity within the spicule group. Green/black pixels in the Doppler/line width maps denote regions too dim to retrieve a quality fit. See also the accompanying animation for this figure. 
\label{fig2}}
\end{center}
\end{figure}

Three parameters (rotation angle, x-center, and y-center) were varied to arrive at alignment parameters which matched both the x-y images (the \ion{Mg}{2}~k at 2796.35~\AA~ and \ion{Ca}{2}~H at 3969.59~\AA~ lines are formed in similar regions of the atmosphere) and time-distance plots using visual inspection. Because of the drift in each data series and our parameterized correction, we estimate the co-alignment of the \textit{IRIS} and SOT datasets is accurate to $0.^{\prime\prime}5$.

\subsection{Spectral Analysis of Spicules}

In Fig.~\ref{fig1} most of the overlapping FOV between \textit{IRIS} and \textit{Hinode}/SOT is shown, in which the \textit{IRIS} slit sits on top of spicules. The yellow square in the figure corresponds to the FOV shown in Fig.~\ref{fig2}. What appears to be a collection of spicules (see arrows in Fig.~\ref{fig2}) at high resolution with SOT in \ion{Ca}{2}~H emanates from a bright region on disk, extending for about 12~Mm with an upwards instantaneous speed of 25~km~s$^{-1}$. Although some sub-structure is also seen with SJI~2796, the collection of strands is mostly seen as a single monolithic structure in that passband. Some substructure is seen in SJI~1400, but the monolithic structure is less clear. As the structure extends upwards along an inclined trajectory, the inner strands are also seen to oscillate in the transverse direction. This sideways motion appears coherent for the set of strands (or spicules) over a transverse length of about 3~Mm, and also along their length, suggesting either a collective standing wave (no upward propagating wavefront is observed) or a long wavelength propagating wave. The oscillation is clearly seen for about one period $\approx250\pm20~$s. During the second half of the oscillation period the collective set of strands brightens simultaneously in all wavelengths. The strand at the right boundary of this group is the brightest and marks the right boundary of the monolithic structure seen in \ion{Mg}{2} (and in \ion{Si}{4}, but less clearly). After this single period motion the whole group fades out in the \ion{Ca}{2}~H passband, but remains visible in \ion{Mg}{2} (and barely in \ion{Si}{4}), a change in visibility that is typical of type II spicules \citep{Skogsrud_2015ApJ...806..170S}. Later on, the spicule group decreases in intensity in all wavelengths and the downward parabolic trajectory is barely visible. Other spicules with different inclination also come into the same LOS, making the follow-up dynamic motion difficult to track.

\begin{figure}[!ht]
\begin{center}
\includegraphics[scale=0.7]{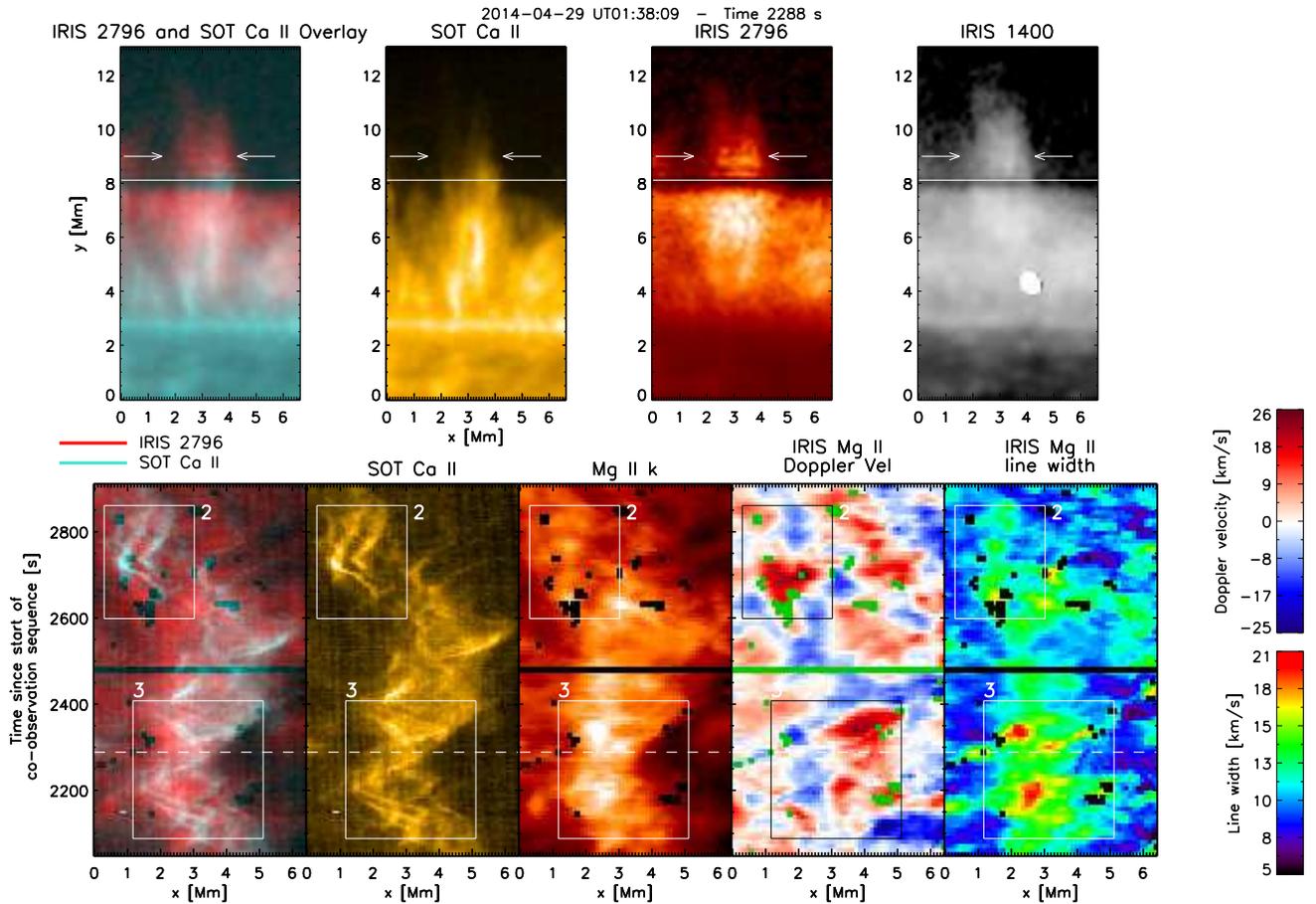}
\caption{\textit{Hinode} and \textit{IRIS} observations of spicules. Similarly to Fig.~\ref{fig2} (case 1), we show here two other cases of spicule groups: one subject to a propagating kink wave (case 2), and another one subject to a rotation or a torsional Alfv\'en wave (case 3). See also the accompanying animation for this figure. 
\label{fig3}}
\end{center}
\end{figure}

Coinciding with the time of both (right- and left-most) maximum transverse displacements, the Doppler shifts change sign with average values between $-10$ and $10~$km~s$^{-1}$ (and maxima around $15~$km~s$^{-1}$). Although the group of strands appears coherent, the Doppler shift transition (sign change) is ragged, being at slightly different times across the width of the  group. The line width across the group shows an enhanced value of $\approx12~$km~s$^{-1}$, about $3-5~$km~s$^{-1}$ above the background. Within the group, the bright right-most strand stands out, with enhanced values up to $19~$km~s$^{-1}$. 

\begin{figure}[!ht]
\begin{center}
\includegraphics[scale=0.7]{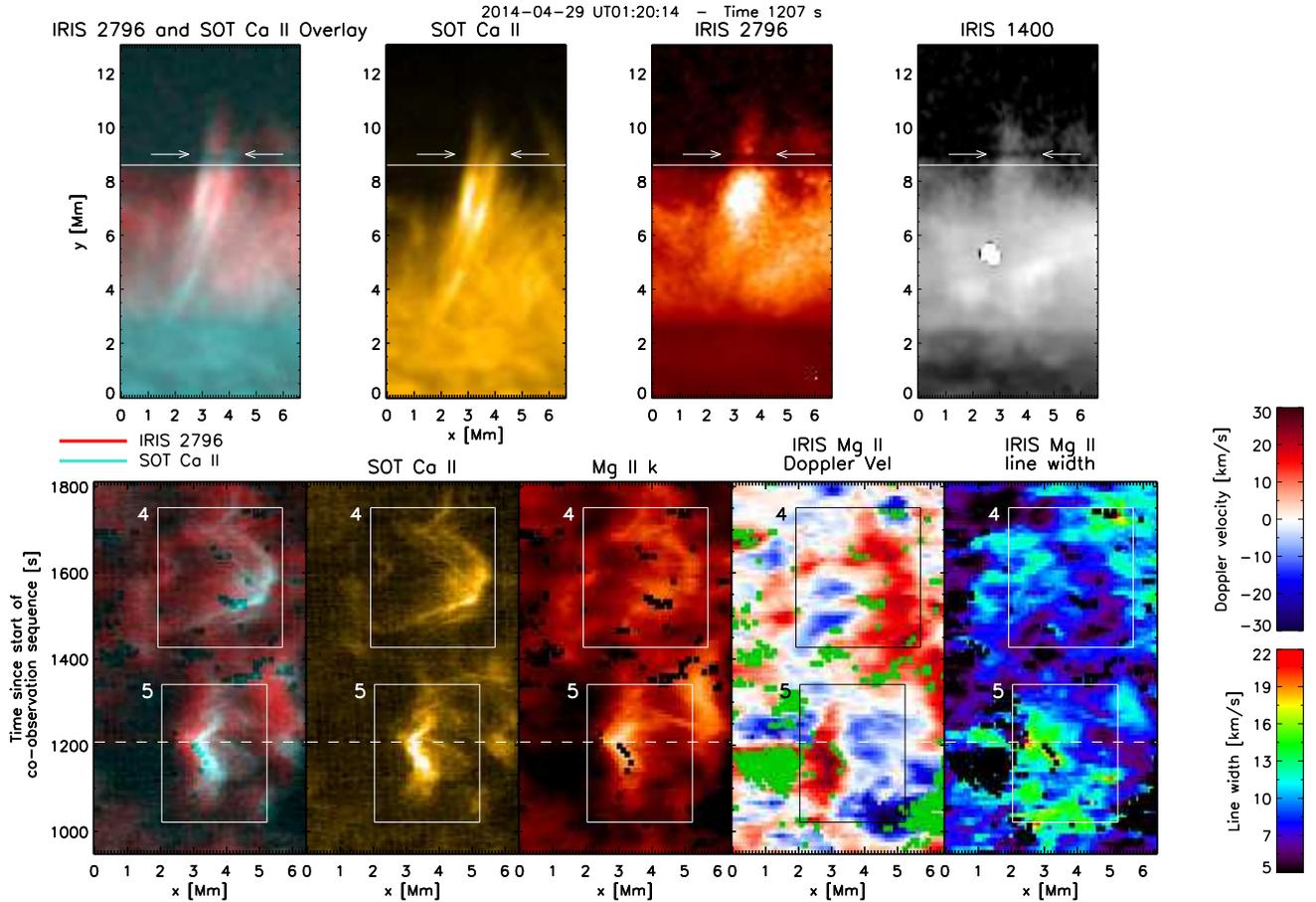}
\caption{\textit{Hinode} and \textit{IRIS} observations of spicules. Similarly to Fig.~\ref{fig2} (case 1), we show here two other cases of spicule groups, subject to a rotation or a torsional Alfv\'en wave (cases 4 and 5). See also the accompanying animation for this figure. 
\label{fig4}}
\end{center}
\end{figure}

In Figs.~\ref{fig3} and \ref{fig4} we show 4 other spicules cases. Case~2 in Fig.~\ref{fig3} shows a similar case as in Fig.~\ref{fig2}, in which, however, an outward propagating transverse displacement is observed (time-distance cuts at different heights show propagating wavefronts; not shown here). The intensity and spectral features appear very similar, albeit shorter lived. 

As for case~1, case~3 in Fig.~\ref{fig3} and cases 4 and 5 in Fig.~\ref{fig4} exhibit strand-like structure and fast disappearance in \ion{Ca}{2}~H, together with persistent, monolithic structure in 2796 and 1400 (with some signatures of strand-like structure). However, the collective motion of the strands is mostly seen out-of-phase and exhibits similar patterns as those reported by \citet{Okamoto_2016ApJ...831..126O} for rotation in a prominence. In particular, the Doppler velocity transitions are oriented roughly along the spicule axis, with one side of the strands group having a distinct blue or red-shifted velocity, opposite to the other side.

These five events illustrate the complexity of multi-wavelength spectroscopic observations of spicules. To make sense of these observations, and particularly the observed difference in the Doppler shifts, it is important to first have a top-level view of what kind of patterns to expect in the data, based on usual agents that are commonly invoked to explain the observed wave-like dynamics, such as transverse MHD waves, torsional Alfv\'en waves, and rotation. We will then first use simple models of these likely scenarios before performing numerical simulations.

\subsection{A Gedanken Model of Spicule Oscillations}

\begin{figure}[!ht]
\begin{center}
\includegraphics[scale=0.4]{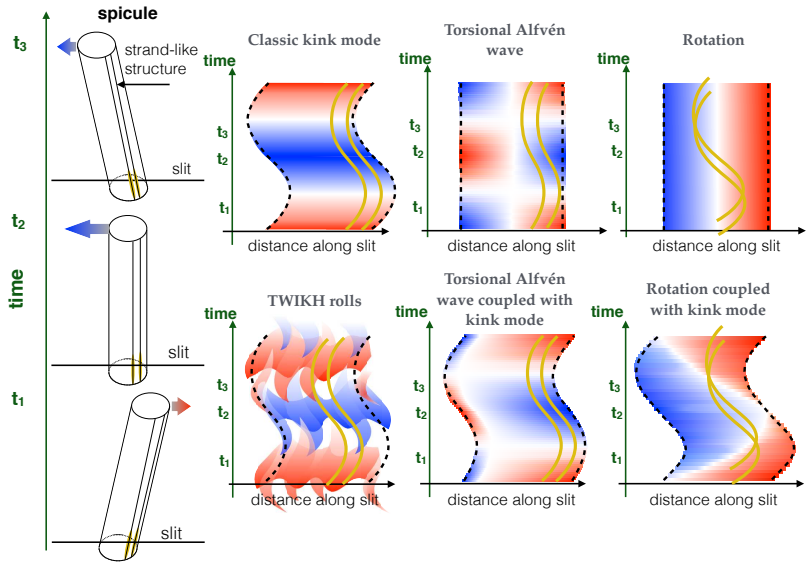}
\caption{Sketch of time-distance diagrams of Doppler velocities for various considered models. The spicule is viewed from the side, as in Fig.~\ref{fig7}. The strand-like structure within the spicule (in yellow) has Doppler signatures (yellow paths) in the time-distance diagrams. In the torsional Alfv\'en wave case we only show a `shell' where the Alfv\'en speed is constant. 
\label{fig5}}
\end{center}
\end{figure}

In Fig.~\ref{fig5} we schematically show the time-distance diagrams of what would be expected from commonly considered scenarios for explaining the observed cases 1 to 5, particularly concerning the distinct Doppler velocities. We consider two kink wave models (the `classic' kink mode and TWIKH rolls model, corresponding respectively, to models with and without associated dynamic instabilities and resonant absorption), a torsional Alfv\'en wave model, a rotation model, and combinations of these two with a kink, assuming that the amplitudes of the torsional and rotational motions dominate. 

Although it is unclear at this point what the strand-like structure would correspond to in these models, the Doppler shifts show a distinct behaviour. The models involving rotation have a Doppler shift transition that is roughly fixed in space, leading to an asymmetry of blue and redshift from left to right of the spicule, potentially matching cases 3, 4 and 5 (cases 3 and 4 seem to combine as well an additional transverse component, such as that produced by a coupling of rotation with a kink mode). This behaviour could also correspond to a propagating long wavelength torsional Alfv\'en wave (assuming we see only half a period or less). The observed Doppler shift transition of cases 1 and 2 is only obtained in the kink mode and partly the torsional Alfv\'en mode. However, for the latter we should observe a periodic asymmetry for half or part of the flux tube, which is not observed. The kink mode, being a collective mode, is particularly well suited for explaining the collective motion of the strand-like structure. On the other hand, to explain the strong collective transverse motion by a torsional Alfv\'en mode, the strands would need to be on a shell of constant Alfv\'en speed within the flux tube, and the wavelength would need to be much longer than the spicule length. Such configuration could be expected to lead to small Doppler shifts due to the sum over the many expected shells with different Alfv\'en speeds within a flux tube (phase mixing, particularly in conditions not far from optically thin), combined with an increase of line width towards the edges, which is not observed. However, in the case of a spectral line having a source function peaking in a shell in the boundary (ring-shaped when taking a cross-section of the loop) a collective strand motion with high Doppler shifts may be possible. Indeed, such a source function would act as a preferential visual filter for the torsional Alfv\'en waves within the shell delimited by the source function. As we will see from our model, such source functions are possible, and only a 3-D MHD simulation of the torsional Alfv\'en wave mechanism, combined with forward modelling, would allow to properly assess this possibility.

In this paper we concentrate on the effect of the kink wave. We will focus particularly on case 1 and check with a 3-D MHD simulation, as the simple cartoon suggests, a flux tube oscillating with a transverse MHD wave can explain the observed Doppler shifts, as well as the other spectral characteristics.

\section{Numerical modelling}\label{model}

As mentioned in the introduction, transverse, wave-like motions are a key property of spicules that cannot be explained by a combination of geometry and their longitudinal up/down motions alone. For the sake of simplicity we assume that we can isolate the transverse motion from the longitudinal motion in order to study the effect that transverse MHD waves have on the observable properties of spicules. In this work we therefore do not focus on the generation mechanism of spicules but assume that we have initially a static spicule at the footpoint of a loop.

\subsection{Numerical model}

\begin{figure}[!ht]
\begin{center}
\includegraphics[scale=0.7]{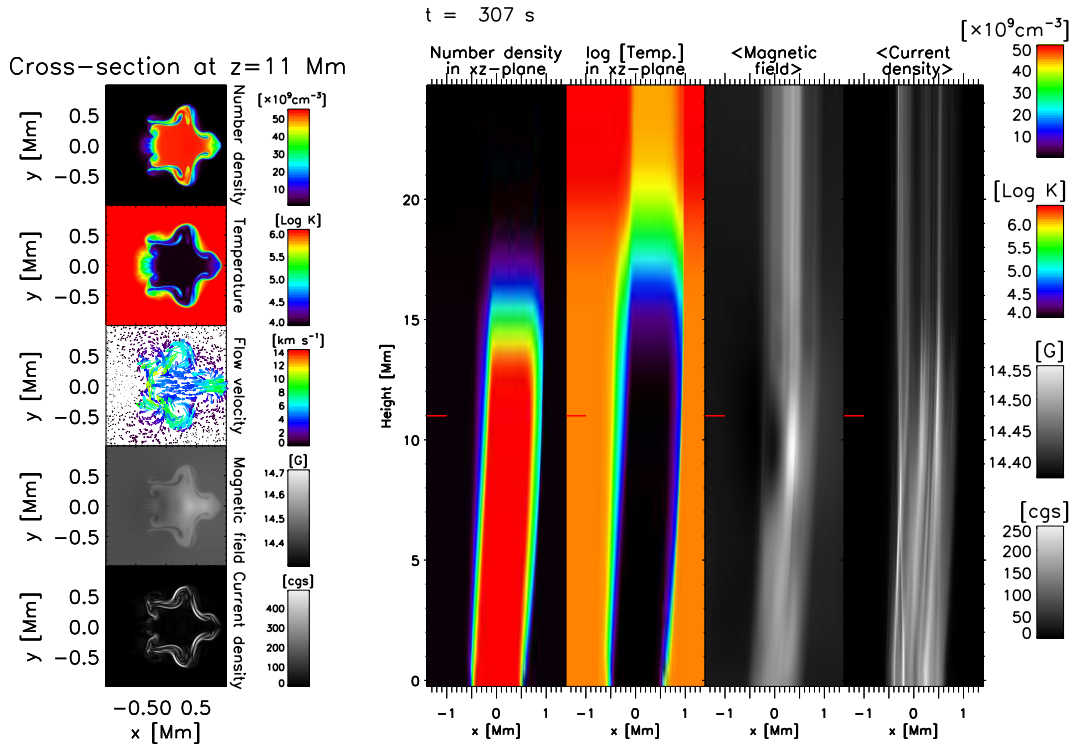}
\caption{Spicule model. \textit{4-set on right:} side view of the simulated spicule along $(x,z)$ plane (plane of oscillation at $y=0$) for the number density and temperature (in log values), and the average along the $y$ axis for each $(x,z)$ coordinate for the magnetic field strength and current density. \textit{5-set on left:} perpendicular cross-section of the spicule at a height of $z=11$~Mm for the same quantities, including the flow velocity. See also the accompanying animation for this figure. 
\label{fig6}}
\end{center}
\end{figure}

As shown in Fig.~\ref{fig6} we model what can be considered a typical spicule \citep{Tsiropoula_2012SSRv..169..181T} with a dense and cold core with $n_i=50~n_e$ and $T_i=\frac{1}{100}T_e$, where $n$ and $T$ denote the total number density and temperature, and $i$, $e$ denote the internal and external values, respectively. We take initially $T_i=10^4~$K, $n_i=6\times10^{10}$~cm$^{-3}$, values commonly found in spicules. The density is given by $\rho=\mu m_p n$, where $\mu$ is the mean molecular weight and $m_p$ is the proton mass. To keep pressure balance throughout the atmosphere, the magnetic field varies slightly from $B_i=14.5~$G to $B_e=14.43$~G, and the plasma $\beta$ is 0.02 everywhere. The boundary layer connecting the internal and external values follows the equation: 
\begin{equation}\label{eq1}
\rho(x,y) = \rho_e+(\rho_i-\rho_e)\zeta(x,y),
\end{equation}
where
\begin{equation}\label{eq2}
\zeta(x,y) = \frac{1}{2}(1-\tanh(b(r(x,y)-\frac{1}{2}))).
\end{equation}
The coordinates $x$ and $y$ are in the plane perpendicular to the loop axis, and $z$ is along the axis. The $r(x,y)=\sqrt{x^2+y^2}/w$ term denotes the normalised distance from the loop centre and $w=1~$Mm denotes the loop width. We therefore assume that our spicule and loop have a circular cross-section with an inhomogeneous boundary layer. We set $b=16$, leading to a boundary layer width of $\approx200~$km. The loop has length $L=100~$Mm.

The upper boundary of the spicule is a transition region to the corona. The internal and external temperatures change everywhere with height according to: 
\begin{equation}
T_{i,e}(z) = T_{i,e}(0)+T_{tr}(1-\tanh(\frac{z-z_{tr}}{w_{tr}})),
\end{equation}
where $T_{i,e}(0)$ are the internal, external temperatures at height $z=0$, $T_{tr}=4.5\times10^{5}$~K, $z_{tr}=15~$Mm and $w_{tr}=2~$Mm are the midpoint and extent of the transition region. The internal and external coronal temperatures are $0.91\times10^6~$K and $1.9\times10^6~$K. The loop has only an extremely mild density enhancement of 1.044 in the corona, leading to a value of $n_i=6.3\times10^{8}$~cm$^{-3}$, and would thus be indistinguishable from the surrounding (diffuse) corona. The initial cross-section and longitudinal variation along the $x-z$ plane can be seen in the animation corresponding to Fig.~\ref{fig6}.

Following case~1, the flux tube is driven with a fundamental kink mode by imposing a transverse perturbation along the loop according to $v_{x}(x,y,z) = v_0 \cos(\pi z/L)\zeta(x,y)$, where $v_0=33~$km~s$^{-1}$ is the initial amplitude (at the loop apex), corresponding to 12~km~s$^{-1}$ in the top part of the spicule. We have chosen the same $(x,y)$ dependence as that of the density in order to maximise the amount of energy given to the fundamental mode, and minimise the generation of other longitudinal modes. The chosen amplitude values are usually found in spicules, and are much lower than the observed maximum speeds \citep[on the order of 60~km~s$^{-1}$,][]{Pereira_2012ApJ...759...18P}. 
The kink phase speeds at spicular level and corona are 256~km~s$^{-1}$ and 1756~km~s$^{-1}$, respectively.

A parameter space investigation was performed for our model, by changing the density (and temperature) contrast, the magnetic field strength, the initial velocity perturbation (amplitude and localisation of the perturbation), and also the shape of our flux tube (multi-stranded structure instead of monolithic). Although various differences are seen (mainly concerning the amplitude of the driver), the physical mechanisms described here remain largely the same.

\subsection{Numerical setup}

The 3-D MHD simulation is performed with the CIP-MOCCT scheme \citep{Kudoh_1999_CFD.8}. The MHD equations exclude gravity, loop expansion and curvature since we focus on the transverse dynamics observed in spicules, for which we assume they are second order factors. Furthermore, the effects of radiative cooling and thermal conduction are not included. The numerical grid contains 512 $\times$ 256 $\times$ 100 points in the $x, y$ and $z$ directions respectively. Thanks to the symmetric properties of the kink mode only half the plane in $y$ and half the loop are modelled (from $z=0$ to $z=50~$Mm), and we set symmetric boundary conditions in all boundary planes except for $x$, where periodic boundary conditions are imposed. To minimise the influence from lateral boundary conditions, the spatial grids in $x$ and $y$ are uniform with a resolution of 7.8~km where the loop dynamics are produced, and increase further away leading to a maximum distance of $\approx 8~$Mm from loop centre. The simulation is close to ideal with no explicit resistivity or viscosity. From a parameter study, we estimate that the effective Reynolds and Lundquist numbers in the code are of the order of $10^4-10^5$ \citep{Antolin_2015ApJ...809...72A,Antolin_2017ApJ...836..219A}. The temporal variation in temperature in our model is therefore mostly due to adiabatic effects. 

\subsection{Spectral synthesis}

We synthesised spectra from the model to compare with observations. Calculations were carried out in two regimes: using the optically thin approximation with the FoMo code \citep{VanDoorsselaere_0.3389/fspas.2016.00004} for the \ion{Si}{4}, \ion{Mg}{2}~k and \ion{Ca}{2}~H lines, and by solving the full non-LTE radiative transfer problem with partial frequency redistribution on a column by column basis with the RH 1.5-D code \citep{Pereira_Uitenbroek_2015AA...574A...3P, Uitenbroek_2001ApJ...557..389U} for the \ion{Ca}{2} and \ion{Mg}{2} lines. Both approaches were done for comparison and for facilitating the calculations (optically thin over radiative transfer) when appropriate. 

The MHD simulation does not explicitly calculate the electron density $n_e$, and to obtain self-consistent values of $n_e$ for the non-LTE radiative transfer calculations we employ a two step process. First, we solved the non-LTE problem for a 5-level plus continuum hydrogen atom, using the Saha-Boltzmann equation to calculate $n_e$. This gives us the non-LTE hydrogen populations and ionisation fraction, which are then used to calculate a consistent $n_e$ in a second run where we solve the non-LTE problem simultaneously for a \ion{Mg}{2} and a \ion{Ca}{2} atom. We use the 5-level plus continuum model \ion{Mg}{2} atom of \citet{Leenaarts_2013ApJ...772...89L}, and a similar 5-level plus continuum \ion{Ca}{2} atom. In both cases we allow for partial redistribution (PRD) in the lines of interest (\ion{Mg}{2}~h \& k, \ion{Ca}{2}~H), and use the fast angle-dependent PRD approximation of \citet{Leenaarts_2012AA...543A.109L}.

The 1.5-D approach was used to avoid the large computational burden of full 3-D calculations. This approximation breaks down near the line cores \citep{Leenaarts_2013ApJ...772...90L}, but because we work with line-integrated quantities or whole line fits, the approximation is likely reasonable. With relatively low column masses in the model, the photon mean free path can be large compared with the cell size. Sharp temperature or density variations can induce large spatial variations in spectra from 1.5-D calculations, when compared with 3-D results that include diffusion from inclined rays. To mitigate for this effect, we spatially convolved the spectra from RH 1.5-D with a Gaussian, and chose a conservative FWHM of 50~km to mimic the typical diffusion in integrated line intensities. This value is based on estimates of the photon mean free path for the \ion{Mg}{2}~h\&k lines in the chromosphere \citep{Leenaarts_2013ApJ...772...89L}. For calculating the source function of the \ion{Mg}{2}~k line we mimic the IRIS filter transmission function by using a Gaussian with $4~\AA$~FWHM and centred at $2796.3~\AA$. For \ion{Ca}{2}~H we use the BFI filter transmission function given by \textit{SSW}.

\section{Results}\label{results}

\subsection{Development of the numerical experiment}

\begin{figure}
\begin{center}
\includegraphics[scale=0.6]{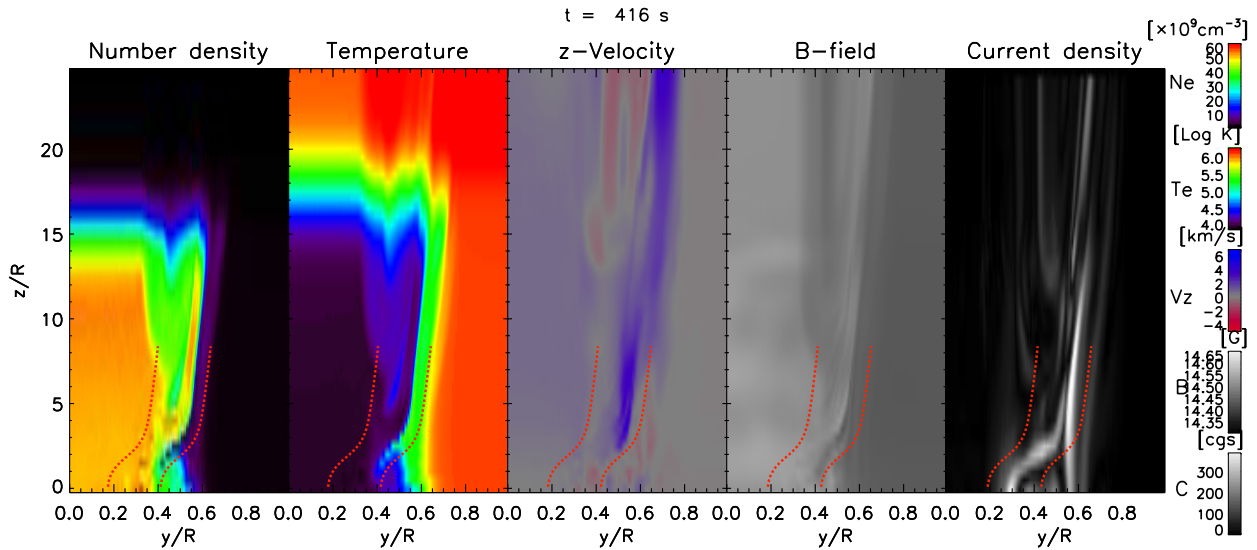}
\caption{From left to right, a cut along the $y-z$ plane at $x=0$ for the number density, temperature, $z-$velocity, magnetic field strength and current density for a particular snapshot in which an upward propagating twist is observed (marked between the red dotted curves). Blue and red colours in the $z-$velocity panel denote upward and downward velocities, respectively. See also the accompanying animation for this figure. 
\label{fig7}}
\end{center}
\end{figure}

The loop oscillates with a period of $\approx255~$s (see Fig.~\ref{fig6}), in agreement with the expected fundamental period (considering the different travel times along the loop) of $2(L_c/c_{k,c}+2L_{ch}/c_{k,ch}+2L_{tr}/c_{k,tr})\approx255$~s, where $L_c$, $L_{ch}$ and $L_{tr}$ denote the lengths of the corona, spicule and transition region, respectively (and $c_{k,tr}$ is an average between the coronal and chromospheric kink speeds). Besides the fundamental mode, higher harmonics and slow and fast propagating modes are produced along the loop. This is particularly evident along the spicule, which sustains its own trapped modes at high frequencies due to the rapidly changing density at the transition region (see magnetic field panel in Fig.~\ref{fig6}).

After half a period we notice the formation of TWIKH rolls all along the loop. The vortices fully develop first in the corona, and about 10-20~s later at chromospheric level. The vortices occur not only at the sides (facing the $y$ direction), but also in the wake of the loop, due to the large velocity amplitude. The vortices at the wake result from the convergence of the azimuthal flow at the wake (see density and vorticity panels in Fig.~\ref{fig9}), where it becomes unstable. The oppositely directed azimuthal flow is mainly driven by azimuthal Alfv\'en waves from resonant absorption and from the inertia of the flux tube, which compresses the front and adds momentum to the flow. Finger-like structures are generated at the wake, which end up curling before moving in the opposite direction. These vortices seem to combine a  Rayleigh-Taylor nature and Alfv\'enic vortex shedding \citep{Gruszecki_2010PhRvL.105e5004G}.

TWIKH rolls twist the magnetic field. Because of the variation of the amplitude of the vortices with height (increasing on average) the twist also varies with height. This differential twist in the longitudinal direction propagates as azimuthal Alfv\'en waves along the loop, non-linearly generating longitudinal ponderomotive forces that drive downward and upward flows. This is shown in Fig.~\ref{fig7}, where an upward propagating twist that has previously reflected at the lower boundary can be seen in the spicule. Although a full representation of this twist would require a follow-up of the 3D geometry of the magnetic field, here we show only a planar cut through the twist in the $y-z$ plane. The $x-y$ cross-section of this twist can be further inspected in Figs.~\ref{fig6} and \ref{fig9} and the accompanying animations. This twist, which constitutes a variation of the magnetic field strength of about 0.2~G (with maxima up to 0.7~G), generates an upward force leading to flows of $5-10$~km~s$^{-1}$. This can also be understood by order of magnitude estimates, since $v_s/v_A\approx B_\phi/B \approx 0.01-0.05$, where $v_s$ is the longitudinal flow generated by the transverse perturbation $B_\phi$ from the twist on the field $B$, and the Alfv\'en speed $v_A$ is on the order of $200-300$~km~s$^{-1}$ at the location of \ion{Mg}{2} formation (within the boundary of the flux tube). The twist is accompanied by currents that will be associated with heating in a non-adiabatic model taking into account resistive and viscous dissipation. While the exact amount of heating would depend on local parameters such as density, the magnetic field strength and resistivity, assuming full dissipation of the Alfv\'enic wave associated with a twist of 0.7~G would lead to an increase of $0.1-1$~MK for a plasma density at the spicule boundary or in the corona of $10^{9}-10^{8}$~cm$^{-3}$, respectively.

A reshuffling of the magnetic field is produced by the TWIKH rolls, which increases and decreases the magnetic pressure around the boundary and towards the core, respectively. Ubiquitous current sheets along the flux tube, throughout most of the loop's volume, are produced in this process. Due to the perpendicular gradient in the magnetic field imposed by the global standing kink mode, both the magnetic field strength and the currents are stronger towards the footpoints than in the rest of the flux tube. This effect, combined with the shearing from the TWIKH rolls, whose size and development increases with height, leads to strong currents that peak at the footpoint and are on average larger along the spicule than in the corona, as can be seen in Fig.~\ref{fig6}. The figure further shows that the KHI vortices introduce not only strand-like structure in density and temperature, but in the magnetic field as well, leading to filamentary current sheets over most of the flux tube.

\subsection{Translating the model into observable quantities}

\subsubsection{Source functions and optical depths}

\begin{figure}[!ht]
\begin{center}
\includegraphics[scale=0.7]{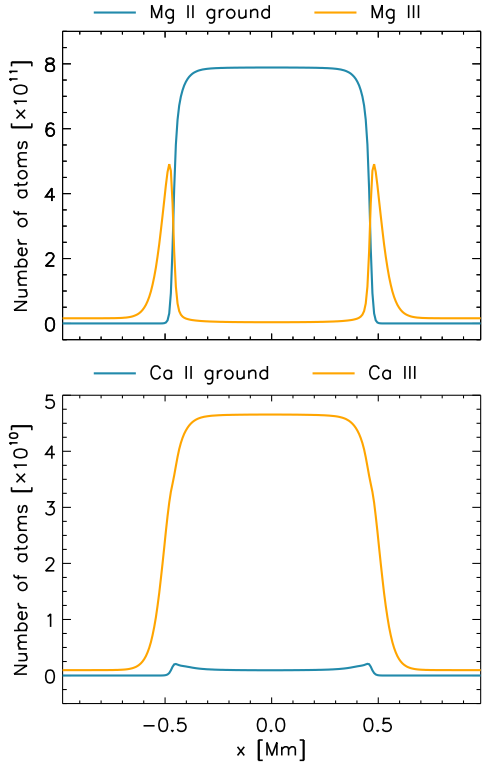}
\caption{Populations of the ground states of the singly ionised and the doubly ionised species for Mg (upper panel), and Ca (lower panel) at the initial time of the simulation (no instability present). 
\label{fig8}}
\end{center}
\end{figure}

\begin{figure}[!ht]
\begin{center}
\includegraphics[scale=0.7]{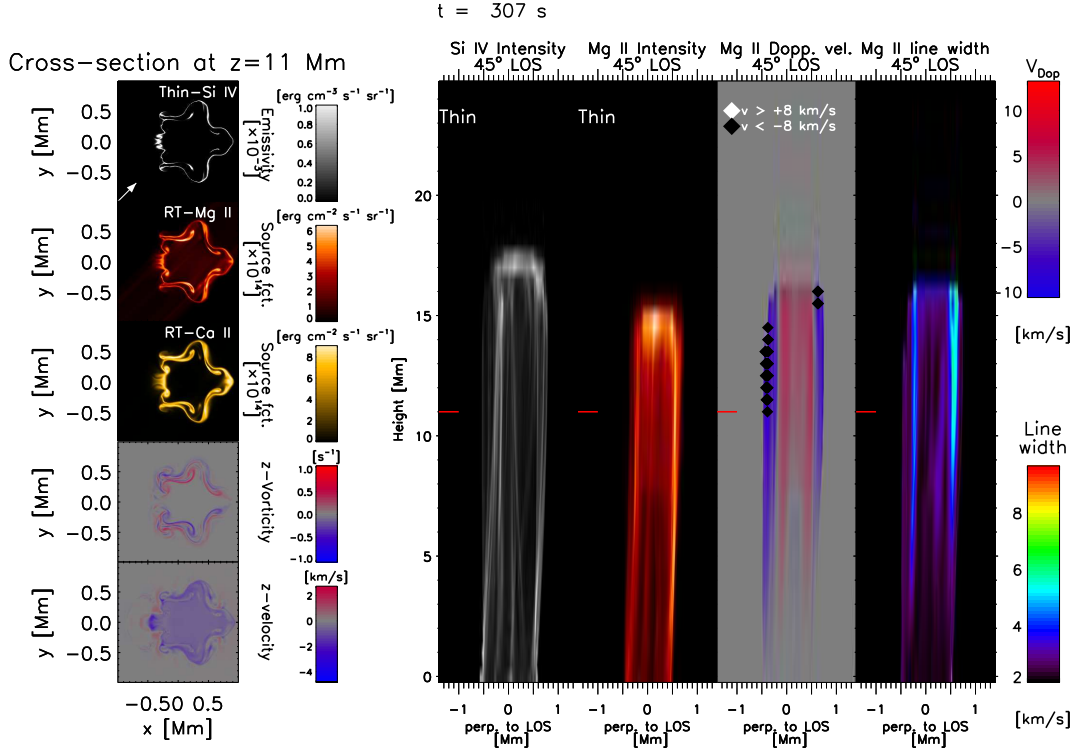}
\caption{Spicule model. \textit{4-set on right:} side view of the simulated spicule at $45^{\circ}$ line-of-sight (LOS) for the optically thin calculation \citep[the LOS angle is defined in the plane perpendicular to the loop, with $0^{\circ}$ matching the $x$ direction, as in][]{Antolin_2017ApJ...836..219A}. The white/black symbols in the Doppler velocity panel indicate velocities above 8~km~s$^{-1}$ in magnitude. \textit{5-set on left:} From top to bottom, perpendicular cross-section of the spicule at a height of $z=11$~Mm for the emissivity ($G_\lambda(T,n_{el})n_{el}^{2}$, with $G, T, n_{el}$ the contribution function for the emission line, temperature and electron number density) of \ion{Si}{4}~1402.77~\AA, the source functions for \ion{Mg}{2}~k and \ion{Ca}{2}~H, the $z$ component of the vorticity and the $z$ component of the velocity. For the latter, blue/red denotes upward/downward velocities. The arrow on the number density panel shows the $45^{\circ}$ LOS direction. 
\label{fig9}}
\end{center}
\end{figure}

\begin{figure}[!ht]
\begin{center}
\begin{tabular}{c c}
\includegraphics[scale=0.4]{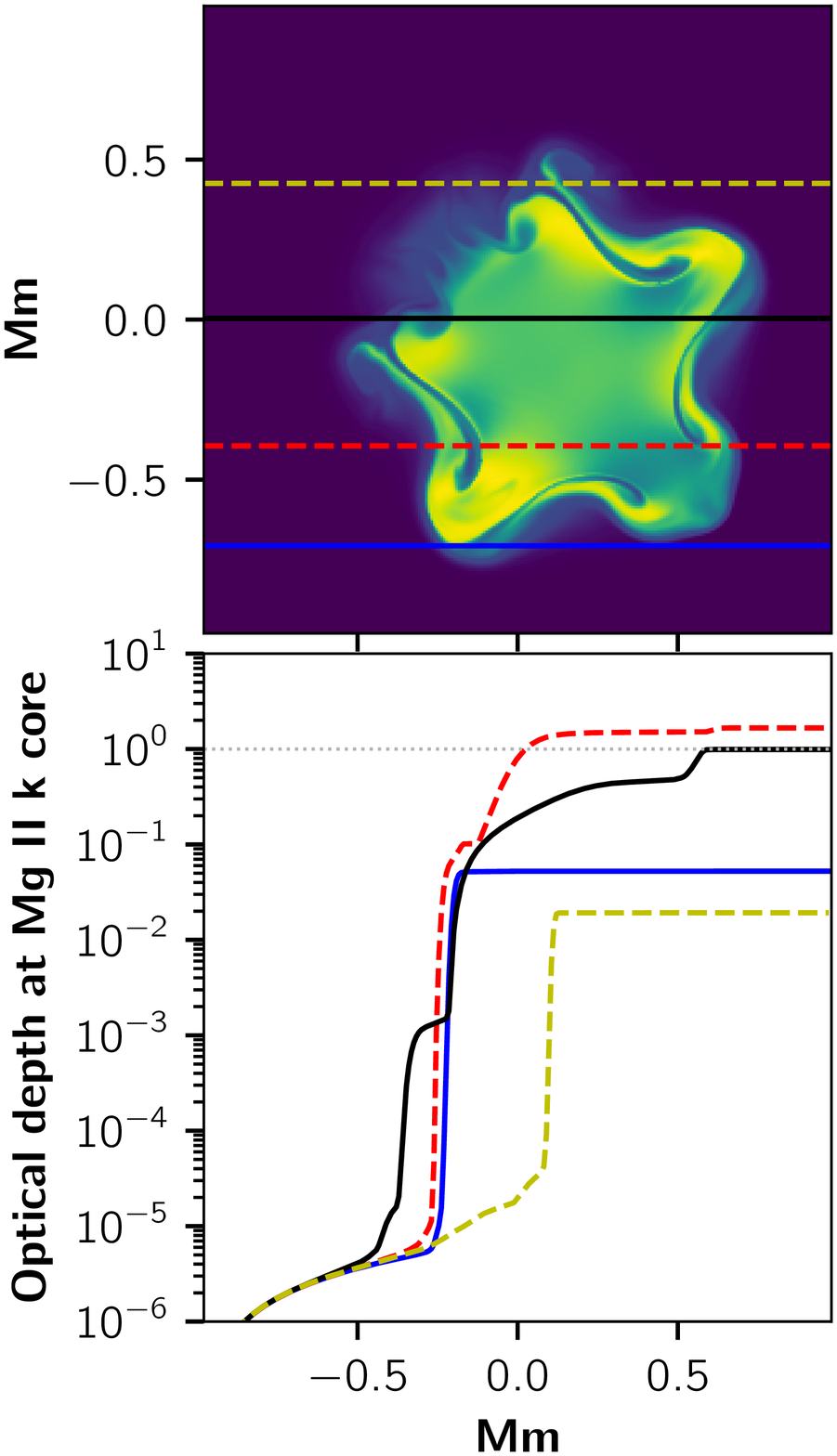} & \includegraphics[scale=0.4]{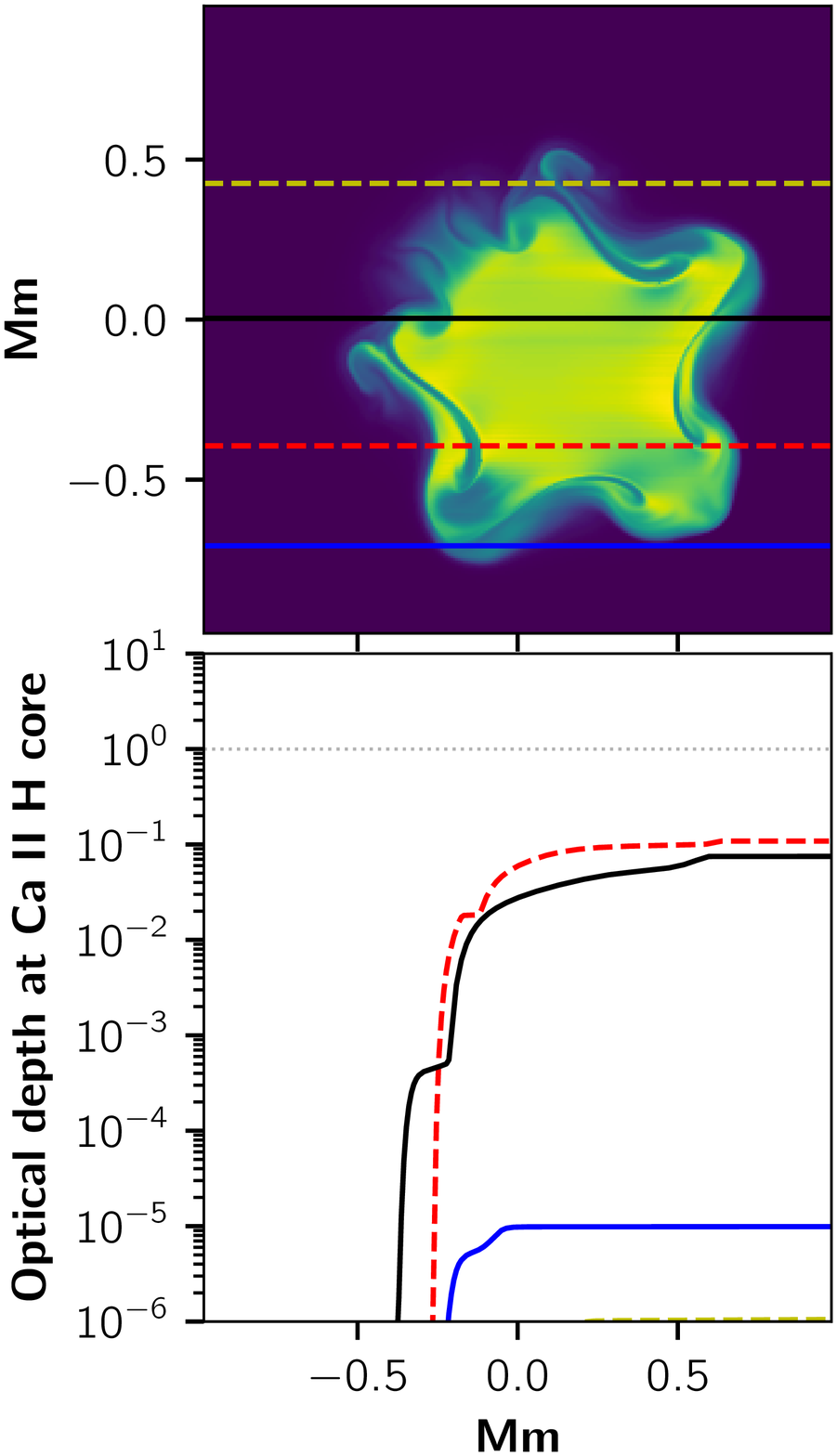} \\ 
\end{tabular}
\caption{Opacity (upper panels) and optical depth (lower panels) for the \ion{Mg}{2}~k (left panels) and \ion{Ca}{2}~H lines (right panels). The solid blue, dashed red, solid black and dashed yellow lines in the upper panels represent LOS rays crossing the loop (background of images correspond to the logarithm of opacity at line centre for each spectral line for the same snapshot as in Fig.~\ref{fig9}). The optical depth for each of these rays at this time in the simulation is shown in the lower panels. Optical depth unity is represented by the dotted grey line.
\label{fig10}}
\end{center}
\end{figure}

Since the lower levels for both the \ion{Mg}{2}~k and \ion{Ca}{2}~H lines are the ground states of their ions, to understand the dynamics observed in these lines we plot in Fig.~\ref{fig8} the level populations of the ground states of singly ionised and doubly ionised species for the Mg and Ca atoms for the initial time. For Mg we find that at the edges of the flux tube there is a sizeable population of \ion{Mg}{3}, but as the temperature drops inside the tube, most Mg is in the form of Mg II, a state that is kept throughout the simulation. This can be seen in the source function cross-section of Fig.~\ref{fig9} -- nearly the whole tube contributes photons to the h \& k lines. The optical depth at line centre for these Mg lines is below 1 for rays going through the boundaries across the KHI vortices, but is on average above 1 for rays crossing through the core of the loop, as shown in Fig.~\ref{fig10}. During the simulation the optical depth varies significantly, and for rays crossing the loop core it can increase to values close to 40. This indicates that the \ion{Mg}{2}~k line is generally optically thick in our model (at the chosen height of 11~Mm). On the other hand, most of Ca in the flux tube exists in the doubly ionised state, \ion{Ca}{3}. Only a few percent of atoms are in \ion{Ca}{2} form, and those are predominantly found in a ring at the edges of the flux tube as shown in Fig.~\ref{fig9}. This region is the only significant contributor to the \ion{Ca}{2}~H line. The optical depth at line centre for rays crossing the loop is well below 1 throughout the simulation. Therefore, the \ion{Ca}{2}~H line is very weak and optically thin.

It is important to note that the described appearance of the spicule in the \ion{Mg}{2}~k and \ion{Ca}{2}~H lines is strongly dependent on the initial condition of our model. For instance, it is likely that a spicule with lower densities will be optically thin in both of these lines, and with different chromospheric temperature ranges it is possible to significantly vary the element population in the core and boundaries, thereby modifying the shape of the respective source functions. In this work we opted for taking values representative of a typical spicule, since the forward modelling of a series of different models that a parameter space investigation would entail, is beyond the present scope, and a very numerically demanding task. We also note that spicules present a highly dynamic environment in which the density may vary on short timescales with lines becoming optically thin and thick during the lifetime of the spicule. Therefore a word of caution is needed for not straightforwardly generalising our forward modelling results.

\subsubsection{Strand-like structure in imaging \& spectra, and propagating twists}

As shown by Fig.~\ref{fig9}, the large TWIKH rolls that appear after one period are observed as opposite large amplitude Doppler shifts and enhanced spectral line widths in the \ion{Mg}{2}~k line at the edges of the loop. This can be understood by the shape of the source function of this line, whose cross-section has a ring structure with peak values at the spicule edges. This, in turn, is due to contributions from wavelengths away from the line core, which contribute to the \ion{Mg}{2} slitjaw emission as observed with IRIS because of the broad IRIS transmission function. At line core the source function is strong in both the edges and inside the tube, but at wavelengths in the line wings the source function is higher at the tube edges because there the regular line profiles are shifted into those wavelengths due to the large Doppler shifts in these regions that come from the KHI dynamics. Therefore, in our model, the \ion{Mg}{2} source function captures well the KHI dynamics, and, at the same time, is shaped by them. As for \ion{Mg}{2}, the \ion{Si}{4} ion has a maximum emissivity around the edge of the vortices, where the temperature increases steeply from chromospheric to coronal values. The \ion{Mg}{2} line has, however, also contribution from the loop's core, leading to a more opaque image compared to \ion{Si}{4}.

The Doppler velocity amplitudes are largest around the times of maximum displacement, where the vortices reach their largest size. The opposite Doppler velocities between the edges and the core are due to the fact that TWIKH rolls have, on average, a similar direction as the counterflow around the flux tube, which is contrary to that of the loop's core (which moves according to the global kink mode).

Downward and upward propagating twists can be seen in the \ion{Si}{4} and \ion{Mg}{2} intensity images of Fig.~\ref{fig9} with speeds of up to 325~km~s$^{-1}$ (the Alfv\'en speed varies between 185 and 1285 km~s$^{-1}$ from the interior to the exterior of the spicule). These twists are associated with the twists identified in Fig.~\ref{fig7}.

As the vortices break up and new vortices appear, strand-like structure is generated along the spicule in intensity, Doppler velocity and line width maps. All quantities shown in Fig.~\ref{fig9} increase with height. For both the \ion{Mg}{2} and \ion{Si}{4} line intensities, this is due to the sharp increase in temperature in the transition region (and corresponding decrease in density), which, under optically thin conditions (which applies to \ion{Mg}{2} at the transition region densities), leads to higher intensity up to the maximum formation temperature ($10^{4.22}$~K and $10^{5}$~K for \ion{Mg}{2}~k and \ion{Si}{4}~1403, respectively).

\subsubsection{Evolution of spectra: bursty profiles for chromospheric lines}

\begin{figure}[!ht]
\begin{center}
\includegraphics[scale=0.7]{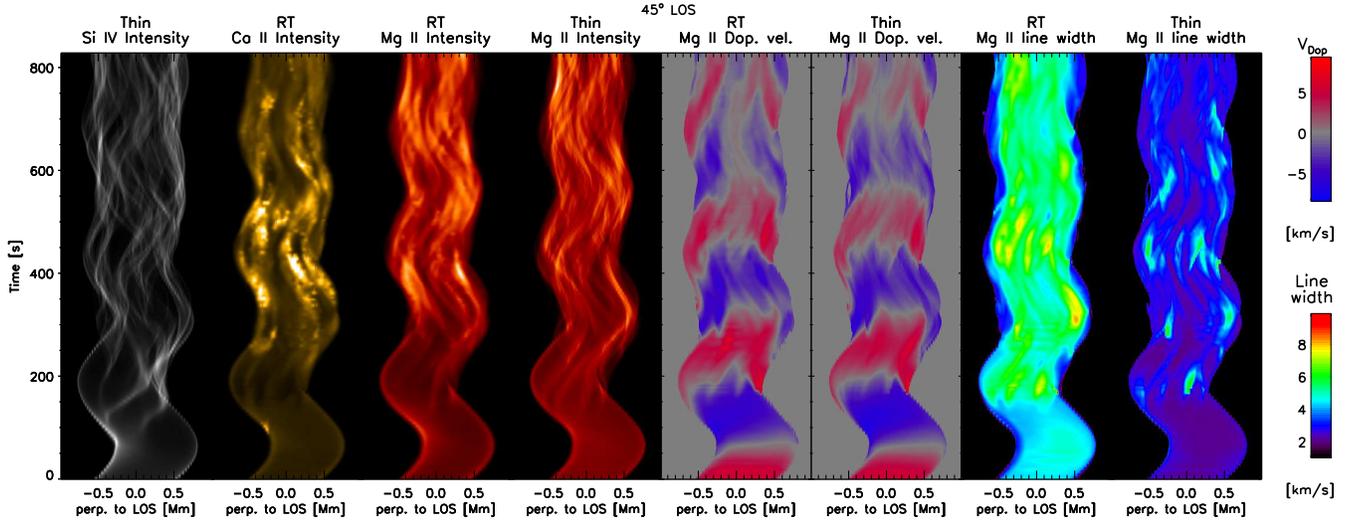}
\caption{Time-distance diagrams at $45^{\circ}$~LOS for the cross-section cut at $z=11~$Mm indicated in Fig.~\ref{fig7} for several quantities (indicated on top) and for optically thin (`Thin') and radiative transfer (`RT') calculations.
\label{fig11}}
\end{center}
\end{figure}

In Fig.~\ref{fig11} we show the time-distance diagram for spectral quantities along a slit crossing  the flux tube perpendicularly at a height of 11~Mm (red line in Fig.~\ref{fig9}). The LOS angle is $45^{\circ}$ with respect to the axis of oscillation (white arrow in Fig.~\ref{fig9}). Results from the optically thin and radiative transfer approaches for the \ion{Mg}{2} line are shown for comparison. The Doppler velocity and line width maps show similar behaviour as previously reported for prominences and coronal loops \citep{Antolin_2015ApJ...809...72A,Antolin_2017ApJ...836..219A}. That is, herring bone shaped features in the Doppler maps, with Doppler shift sign changes starting at the loop edges, maximum values also at the loop edges \cite[$\pi$ out-of-phase with the plane-of-sky motion, as previously observed with \textit{Hinode} \& \textit{IRIS},][]{Okamoto_2015ApJ...809...71O} and ragged transitions between consecutive blue and red shifts due to the TWIKH rolls. A sharp increase in line width (4 to 7 km~s$^{-1}$) is observed accompanying the formation of the vortices. 

 The \ion{Si}{4}, \ion{Ca}{2} and \ion{Mg}{2} intensities show strand-like structures due to the TWIKH rolls, although the loop average intensity and the intensity variations are different for each. Due to the difference in opacity, the loop's body appears mostly invisible in \ion{Si}{4} and similarly in \ion{Ca}{2}, but is visible in \ion{Mg}{2}, which is mostly optically thick. The TWIKH rolls intensity over time is rather constant in \ion{Si}{4}, with occasional bursts on a time-scale of $\approx50~$s. However, the \ion{Mg}{2}, and particularly the \ion{Ca}{2} intensity is strongly bursty with bright features lasting $\approx50-100~$s. 

In our model we find that the Doppler velocities in \ion{Mg}{2} derived from the optically thin approximation generally agree with those derived from the radiative transfer approach, despite the presence of reversals in the line profile. This is mostly due to the ring shape of the source function, which indicates that most of the line profile is formed in the edge (and therefore optically thin) part of the flux tube, even for a LOS ray that crosses the loop core. Since the pixels on the opposite side of the loop along this LOS ray have, in general, a similar Doppler velocity (due to the kink mode symmetry), the contribution of these pixels (present in the optically thin approximation) does not significantly alter the centroid of the line. On the other hand, the line widths do differ between both approaches due to the presence of opacity broadening. The line widths from the radiative transfer calculations are in general larger than those from optically thin radiation by $2-3$~km~s$^{-1}$ (50-60\%). The line profiles obtained from IRIS for case~1 exhibit on average few reversals, suggesting a spicule that is not as optically thick as the one considered in our model.

\section{Discussion}\label{discussion}

\subsection{Detailed comparison with IRIS}

To bring our numerical results closer to the observations we degrade the spatial, temporal and spectral resolutions to those corresponding to the \textit{IRIS} and \textit{Hinode}/SOT observations. For the \textit{IRIS} \ion{Mg}{2}~k line comparison we convolve the data spatially with the \textit{IRIS} PSF. We then convolve the spectra with a Gaussian with FWHM of 60 m\AA~(approximate NUV resolution) and rebin to \textit{IRIS} spectral pixel size. This is followed by spatial pixelisation to the same platescale as IRIS. Lastly, we perform Gaussian fits to the spectra to obtain Doppler velocities and line widths. For comparison with observations with the SOT \ion{Ca}{2} filter we integrate the \ion{Ca}{2}~H spectra over wavelength and apply the spatial convolution using a PSF with FWHM of $0.^{\prime\prime}22$, after which we rebin to the SOT platescale. For the temporal domain, we sum the numerical quantities for a number of snapshots matching roughly the time cadence of the instrument. Since the SOT exposure time is very short (similar to that of the numerical output), we only do the temporal integration for the \textit{IRIS} forward modelling. 

\begin{figure}[!ht]
\begin{center}
\includegraphics[scale=0.7]{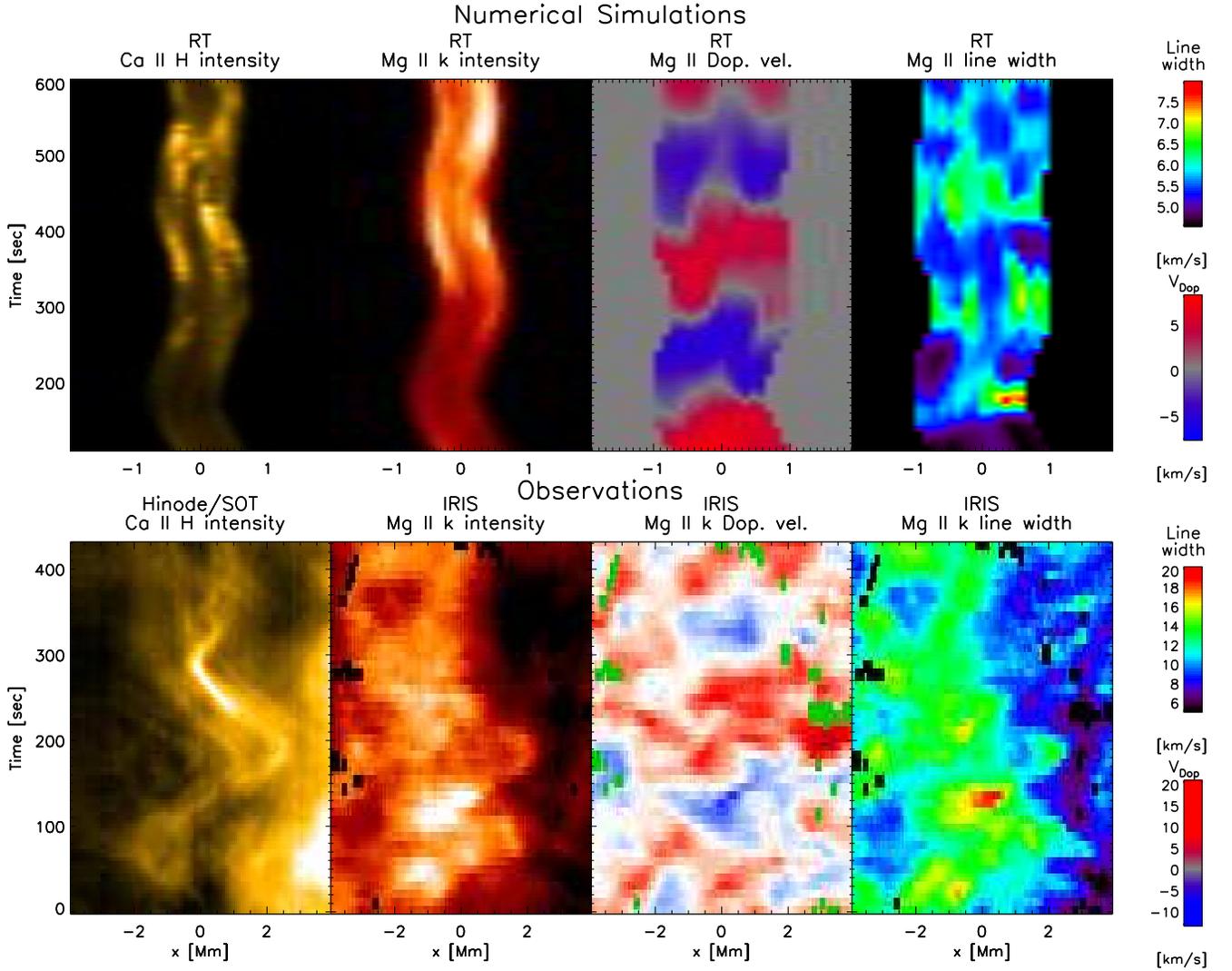}
\caption{Comparison of \textit{Hinode}/\textit{IRIS} observations with simulation results. We forward model the numerical results with a radiative transfer approach and degrade the resolution targeting the \textit{IRIS} and \textit{Hinode} spatial, temporal and spectral resolutions. For the sake of comparison, we take the same temporal interval (in terms of wave oscillation periods) as in the observations and change the sign of the Doppler velocities, which would correspond to a viewing angle of 135$^\circ$.
\label{fig12}}
\end{center}
\end{figure}

In Fig. ~\ref{fig12} we show the results of the forward modelling specified above. For better comparison we take an interval of time in the simulation similar to that observed (particularly in terms of wave periods, which is roughly 2), and show $\approx100$~s into the simulation, which can be interpreted as the time taken by the spicule to move upwards, above the chromosphere (and become visible), and also the time since the start of the kink perturbation \citep[which could be produced by the spicule generation mechanism itself,][]{Martinez-Sykora_2017}. Furthermore, we assume that we are observing from an angle of 135$^{\circ}$ instead of the previously assumed $45^{\circ}$. This is only to match the observed alternation of Doppler shifts, starting from red, for case~1. Because of the symmetries of the kink mode, we expect minimal changes to the radiative transfer modelling from this change of angle. Comparing the numerical results with the observations in Fig.~\ref{fig12} we can better distinguish the similarities and differences. The first obvious difference is the amplitude. Our modelled perturbation leads to roughly half (or less) the observed displacement. The obtained Doppler velocities and line widths are also about half those observed. We would expect that for higher kink amplitudes in our model correspondingly higher velocities and line widths in the TWIKH rolls are obtained. There is also much more complexity in the \ion{Ca}{2} and \ion{Mg}{2} intensities and spectra from the observations, which seems to lead to a clear imbalance in the line widths between the left and right side of the spicule. Besides possible LOS interference from other structures, we also notice that the spicule group is about double the size of our model. The increase in size would, on one hand, be comparable to increase the numerical resolution (if keeping the same number of grid points), which, as shown in \citet{Antolin_2015ApJ...809...72A} and \citet{Magyar_2016ApJ...823...82M}, brings in a myriad of additional vortices of multiple sizes (vortices within vortices recalling a fractal geometry). On the other hand, a larger size also means a larger velocity shear at the boundaries for the same perturbation amplitude, which would more readily trigger the KHI. These effects would therefore increase the complexity in the imaging and spectral quantities and bring the model closer to observations.

Nonetheless, several of the observed features of case~1 in Fig.~\ref{fig2} and case~2 in Fig.~\ref{fig3} (and previously reported) with \textit{Hinode} \& \textit{IRIS} of spicules can be reproduced with the model: the strand-like structure, the collective behaviour among the strands, the (qualitative) variation in \ion{Ca}{2} and \ion{Mg}{2} intensities, the ragged Doppler shift transitions and, to a lower extent, the accompanying line width increase. The collective behaviour observed at high resolution in \ion{Ca}{2}~H (in Fig.~\ref{fig2}) among the strand-like structure opens an interesting question on the nature of the spicule. Is it the single strand structure or the set of strands, seen as a monolithic structure in the coarser resolution of \textit{IRIS} and at higher opacity in the \ion{Mg}{2} line? The collective motion suggests the latter, belonging to a flux tube entity which is mostly invisible in the \ion{Ca}{2}~H line \citep{Skogsrud_2015ApJ...806..170S}. This is supported by the possibility that many of the observed strands in intensity and in Doppler maps may actually be TWIKH rolls, as our model suggests.

\subsection{Higher resolution observations: stranded structure in spectra}

\begin{figure}[!ht]
\begin{center}
\includegraphics[scale=0.7]{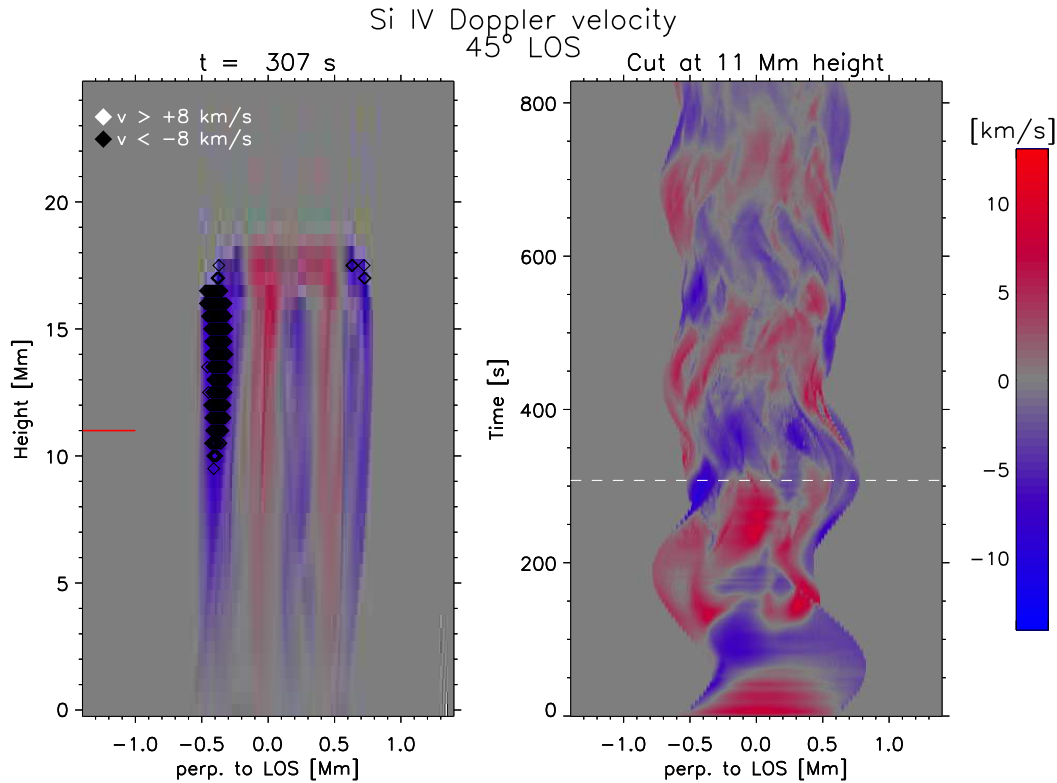}
\caption{Doppler velocity in the \ion{Si}{4} line for a LOS of 45$^{\circ}$. In the left panel the Doppler velocity map is shown for the same snapshot as in Fig. ~\ref{fig9}, where the black (and white) diamonds indicate speeds above 8~km~s$^{-1}$ in magnitude. In the right panel we show the time-distance diagram for a cut at a height of 11~Mm (red line on the left panel). 
\label{fig13}}
\end{center}
\end{figure}

Higher resolution observations with CRISP at the \textit{SST} indicate sudden appearance/disappearance and short-lived strand-like structure in the Doppler shifts across spicule-like features, even with opposite signs for adjacent strands \citep{Pereira_2016ApJ...824...65P, Kuridze_2015ApJ...802...26K,Shetye_2017arXiv170310968S,Srivastava_2017NatSR...743147S}. Such features are perhaps the observed features that are best reproduced from this model once the KHI develops after roughly one wave period time (Figs.~\ref{fig9} and \ref{fig11}). This strand-like structure in Doppler is even more distinct in the \ion{Si}{4} line, as shown in Fig.~\ref{fig13}. In our model, the \ion{Si}{4} line maps a slightly outer layer of the flux tube, and therefore one in which the azimuthal Alfv\'en waves travel at faster speeds and the Doppler shifts have higher frequency. Doppler shift variations with a period of $\approx50-100$~s can be seen at small scales. At the coarser resolution of \textit{IRIS} (Fig.~\ref{fig12}) the strand-like structure in the Doppler maps is not observed, and only a ragged transition in time between opposite Doppler shifts remains, matching well the observed Doppler shift change in case~1. The TWIKH rolls develop after about one period, in agreement with the observed time interval in Fig.~\ref{fig2} from the moment the oscillating spicule appears until the time of maximum strand formation and intensity. 

\subsection{Bursty, doubly periodic line profiles}

\begin{figure}[!ht]
\begin{center}
\includegraphics[scale=0.7]{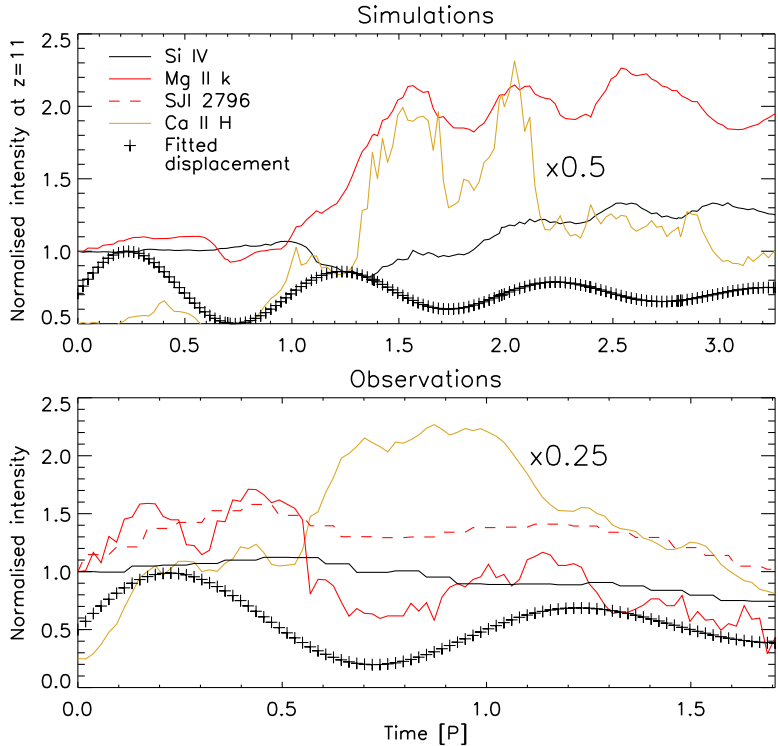}
\caption{Light curves integrated over the spicule width for the \ion{Ca}{2}~H, \ion{Mg}{2}~k and \ion{Si}{4} lines for the simulation (upper panel) and observations (lower panel, we show both the \ion{Mg}{2}~k and SJI 2796 light curves, besides the SJI 1400 light curve from IRIS). To reduce the contribution from other features along the LOS, for the SOT and SJI quantities we integrate also in the vertical direction, from the slit position up a height of 4~Mm). The SOT light curve values are multiplied by a half and a quarter in the top and bottom panels, respectively, for better comparison with the other light curves. The x-axis represents time in units of the wave period, 255~s and 250~s, in the model and observations, respectively. An exponentially decaying cosine has been fitted to the observed SOT displacement in the time-distance diagrams (fitting the maximum intensity crosses in Fig.~\ref{fig2}) and also to the time-distance diagram of the \ion{Mg}{2} intensity map from the simulation. The fits are overlaid (with ad-hoc amplitudes) as crosses for each time step. 
\label{fig14}}
\end{center}
\end{figure}

To further inspect the similarities and differences in the temporal variation of intensity we integrate over the spicule width the \ion{Ca}{2}, \ion{Mg}{2} and \ion{Si}{4} intensities, and show the resulting light curves in Fig.~\ref{fig14}. For the observations, besides the integrated \ion{Mg}{2}~k intensity at the slit position, we also show the \ion{Ca}{2}~H, SJI 2796 \& 1400 filter intensities averaged along the spicule from the slit position to a height of 4~Mm. This is done in order to reduce  the possible LOS interference of other structures at lower heights as well as small-scale longitudinal inhomogeneities, and allows to focus on effects that have a long scale height (such as the global kink mode and the TWIKH rolls). From Fig.~\ref{fig14} we can see that the integrated \ion{Mg}{2}~k intensity has stronger variability than the SJI~2796 intensity, which is expected because of the single slit position, but both lines follow roughly the same trend and are roughly in-phase with the spicule's transverse displacement. The fact that both lines follow the same trend is a reassurance that most of the captured signal at the IRIS slit position in the observations comes indeed from the spicule of interest, and not from other structures along the LOS.

To some extent, the strong variability in \ion{Ca}{2}~H, matching previously reported appearance and disappearance of strands in this line, and the accompanying increase in \ion{Mg}{2}~k intensity (and SJI 2796), are well reproduced in the numerical model. The SJI~1400 light curve shows very little increase for case~1, and a decrease at the end of the time series is observed in all light curves, corresponding to a general fade out during the downward motion of the spicule. This small intensity increase in \ion{Si}{4} seems to be captured by the numerical model. In our model we observe a clear in-phase intensity increase for all light curves, with a period half that of the kink mode and with an offset of roughly $\frac{\pi}{2}$ from the maximum transverse displacement of the spicule, coinciding with times of maximum velocity shear. This offset is associated with the formation of the TWIKH rolls \citep{Antolin_2017ApJ...836..219A}. However, in case~1 the in-phase behaviour between light curves is not clear. While initially a similar trend is observed between \ion{Mg}{2}, \ion{Ca}{2}~H and \ion{Si}{4}, the subsequent strong increase in \ion{Ca}{2} seems to break the in-phase behaviour. The presence of periodicity with a period half that of the kink mode is also unclear.

\subsection{Heating or just mixing?}

In most reported cases an increase in SJI 1400 intensity is found, following the increase in cooler chromospheric lines, unlike the observed very mild increase of case~1 \citep{Skogsrud_2015ApJ...806..170S}. Indeed, previous and recent work indicates a transition region and coronal response to spicules, observed not only as an increase in \ion{Si}{4} emission but also in the coronal filters of AIA, suggesting heating \citep{DePontieu_etal_2011Sci...331...55D, DePontieu_2017ApJ...845L..18D}. These results have been explained from a 2.5-D radiative MHD model, in which similar features are obtained from the combination of propagating currents and heating fronts \citep{DePontieu_2017ApJ...849L...7D}. In these models tangled fields are generated in the photosphere or below and ambipolar diffusion allows those fields, normally too weak to emerge into the atmosphere, to emerge and release the magnetic tension. High amplitude Alfv\'enic waves are produced, which non-linearly generate strong enough slow mode shocks that drive material upwards thereby forming a spicule. These Alfv\'enic perturbations propagate along the spicule and corona, heating up the plasma. The combination of slow mode shocks, longitudinal flows an Alfv\'enic perturbations lead to coronal propagating disturbances and generation of coronal strands that are similar to those observed. Although the resistivity in the spicule and chromosphere are captured well by the ambipolar resistivity in that model (which is higher than the numerical resistivity in the spicule), the obtained Alfv\'enic wave dissipation in the coronal parts of the model is strongly dependent on the resistivity, which is much higher than the expected coronal resistivity. The model presented here shows that Alfv\'enic waves such as transverse MHD waves can further lead to TWIKH rolls.

Our model is purely adiabatic, and therefore there is no heat exchange between adjacent fluid elements. The obtained intensity variability is due to compression and mixing from the TWIKH rolls of inner (colder) with outer (hotter) material. Our results therefore require caution when interpreting observations, since we can obtain similar intensity enhancements in chromospheric lines which are not caused by actual heating. While a time-lag in intensity increase is usually observed from chromospheric to transition region lines (interpreted as heating), the TWIKH rolls in our model lead to in-phase variation of the intensity between chromospheric and transition region lines. TWIKH rolls are compressive and locally enhance the plasma density adiabatically. On average, the rolls enhance the magnetic field around the boundary layer, which is at the same time decreased within the flux tube. Correspondingly, the internal energy is decreased around the boundary layer and increased within the flux tube. The obtained variation of intensity, and particularly its increase within a TWIKH roll does not strictly correspond to heating from wave dissipation \citep{Antolin_2017ApJ...836..219A}. Our model however shows that TWIKH rolls enhance the currents through the bending, reshuffling and twisting of magnetic field lines, suggesting that such TWIKH rolls can lead to heating. Indeed, in similar but more realistic resistive (and viscous) models \citep{Karampelas_2017AA...604A.130K}, and resistive but higher resolution \citep[leading to well developed turbulence][]{Antolin_2015ApJ...809...72A} or with anomalous resistivity \citep[simulating energy deposition at smaller scales due to turbulence,][]{Howson_2017AA...602A..74H} wave dissipation is achieved. It is, however, still unclear whether significant heating would take place to counter radiative and conduction losses. The addition of twist in the flux tube as initial condition, even if small, although suppressing to some extent the TWIKH rolls \citep{Soler_2010ApJ...712..875S, Terradas_2017arXiv171206955T}, would also produce more energetic ones, due to the addition of an azimuthal component in the field \citep{Howson_2017AA...607A..77H}. The TWIKH rolls from our model or one with twist (in which case they would be helical) may also explain the vortex motions observed around chromospheric jets \citep{Kuridze_2016ApJ...830..133K}. The higher energies in the twisted scenario would likely lead to stronger field aligned flows \citep{Yokoi_2016PhRvE..93c3125Y}. We have shown that the twists and currents that accompany the TWIKH rolls in our model generate Alfv\'enic propagating disturbances along the spicule with significant energy. It is unclear, however, whether such disturbances would be able to heat the corona in a more realistic model. In an extended 3D MHD model from that of \citet{Martinez-Sykora_2017} but with higher spatial resolution similar to that in our model we would expect a combination of the described effects. Our model suggests that TWIKH rolls would be produced by the transverse perturbations that accompany the spicule generation mechanism. This combination would therefore not only explain the spicule generation but also its multi-strand morphology in intensity and spectra, and further support the accompanying heating at transition region and coronal temperatures from that model.

\subsection{TWIKH rolls in Coronal Loops, Prominences and Spicules}

Besides spicules as in the present work, the TWIKH rolls model has been tested in other very different plasma environments of the solar atmosphere: prominences \citep{Antolin_2015ApJ...809...72A} and coronal loops \cite{Antolin_2014ApJ...787L..22A,Antolin_2017ApJ...836..219A}. The major parameter differentiating each model is the density contrast: $\rho_i/\rho_e=50,10,3$, respectively for the spicule, prominence and the coronal loop. The onset time of the TWIKH rolls increases with the density and the velocity shear amplitude, as expected from theory. The higher the density of the structure, the higher the inertia and subsequent deformation of the flux tube. In the prominence and particularly in the spicule cases, we observe, additionally to the KHI, rolls that develop at the wake displaying characteristic Rayleigh-Taylor features, which help in the subsequent generation of turbulence. 

In all three cases TWIKH rolls play a major role in the appearance of the structure in EUV, transition region or chromospheric wavelengths, with the characteristic strand-like structure. Due to the higher phase speeds at the edges of the loop (leading to phase mixing) the Doppler maps display alternating blue/redshift herring bone shapes, and the TWIKH rolls appear as red/blue shifted intrusions in the main blue/red shifted signal of the flux tube's body. These intrusions are visible at high resolution, and the forward modelling performed here shows that they should be observable at IRIS resolution, offering an explanation for the small-scale features in Case~1.

Higher resolution allows more wave dissipation in our numerical model, due to better developed turbulence \citep{Antolin_2015ApJ...809...72A}. Setting the spatial resolution factor aside, we have shown that the mixing can play an important role in the observed morphology. Since this mechanism depends on the temperatures and densities present initially, the end result can be significantly different. This is illustrated by the coronal, prominence and spicule models (comparing only those at same spatial resolution): for similar perturbation amplitudes and for the same temperature contrast, the population of the `mixed' material will have lower intermediate temperatures for higher density contrast. For instance, we would expect a  higher \ion{Si}{4} population at the end state by taking a lower density contrast in the spicule model (as indicated by the prominence model).

\section{Conclusions}

Having analysed \textit{Hinode} \& \textit{IRIS} observations of several spicules, we found significant differences between them: cases with a fixed Doppler shift transition along the spicule axis and little coherence of strands match best rotation or propagating long wavelength torsional Alfv\'en wave models. Cases with a Doppler shift transition at maximum displacement and strong strand coherence match best the kink wave model. The latter is confirmed with a 3-D MHD numerical and radiative transfer forward modelling of a spicule oscillating with a transverse MHD wave. TWIKH rolls are rapidly produced and can explain several of the observed features: coherent strand-like structure in intensity, Doppler velocities and line widths at high resolution, ragged Doppler shift sign changes at maximum displacement accompanied with increased line widths. The rapid temperature variations from the mixing and the TWIKH rolls short lifetimes produce strong variations in Doppler velocity and in the \ion{Ca}{2}~H intensity, in agreement with the observed sudden appearance and disappearance of spicule strands in chromospheric lines, but not matching the clear preference for fading in cooler lines and appearance in hotter lines. This model can explain particularly well the Doppler maps from high resolution instruments with tunable wavelength filters such as  \textit{SST}/CRISP, where spicules show up as contiguous opposite Doppler shifts (strand-like structure in Doppler maps) that appear and disappear at high frequency. Rather than the individual strand being called a spicule, our model suggests that the collection of such strand-like structure should be considered as one spicule, belonging to the same flux tube that is subject to a transverse MHD mode. The TWIKH rolls also lead to twists and currents that propagate along the spicule at Alfv\'enic speeds. The increase in temperature in our model only results from adiabatic effects, which are unable to generate the observed intensity increase in higher temperature lines, suggesting that an additional mechanism must be at work. This is not surprising given the simplicity of our model, and results from the more realistic 2.5-D MHD model of \citet{Martinez-Sykora_2017} indicate already that spicule-like features are likely to involve a variety of complex mechanisms that are not included here. However, our higher resolution model indicates that the small scale processes triggered by dynamic instabilities associated with transverse MHD waves may also play a significant role in the observed spicule features.

\acknowledgments
We would like to thank the anonymous referee for valuable comments and suggestions that have led to an improvement of this manuscript. This research has received funding from the UK Science and Technology Facilities Council (Consolidated Grant ST/K000950/1) and the European Union Horizon 2020 research and innovation programme (grant agreement No. 647214), and also from JSPS KAKENHI Grant Numbers 25220703 (PI: S. Tsuneta). T. P. acknowledges support by the Research Council of Norway through its Centres of Excellence scheme, project number 262622, and through grants of computing time from the Programme for Supercomputing. Numerical computations were carried out on Cray XC30 at the Center for Computational Astrophysics, NAOJ, and at the Pleiades cluster through computing project s1061 from NASA's High-End Computing Program.

\bibliographystyle{aasjournal}
\bibliography{ms_spic_accepted.bbl}  

\end{document}